\documentclass[12pt, draftclsnofoot, onecolumn]{IEEEtran}
\ifCLASSINFOpdf

\else

\fi
\usepackage[cmex10]{amsmath}
\usepackage{amssymb}
\usepackage{cite}
\usepackage{graphicx}
\usepackage{array,color}
\usepackage{amsmath}
\usepackage{stfloats}
\usepackage{graphicx}
\usepackage{subfigure}
\usepackage{tabularx}
\usepackage{epsfig,epsf,color,balance,cite}
\usepackage{algorithmic}
\usepackage{algorithm}
\usepackage{bm}
\begin{document}
%
\title{Pricing-based Distributed Energy-Efficient Beamforming for MISO Interference Channels}



\author{Cunhua Pan, Wei Xu, \IEEEmembership{Senior Member, IEEE}, Jiangzhou Wang, \IEEEmembership{Senior Member, IEEE}, 
Hong Ren, Wence Zhang, Nuo Huang and Ming Chen
\thanks{C. Pan, W. Xu, H. Ren, W. Zhang, N. Huang and M. Chen are with National Mobile Communications Research Laboratory, Southeast University, Nanjing 210096, China. (Email:\{cunhuapan, wxu, renhong, wencezhang, huangnuo, chenming\}@seu.edu.cn).}

\thanks{J. Wang is with the School of Engineering and Digital Arts, University
of Kent, Canterbury, Kent, CT2 7NZ, U.K. (Email:{j.z.wang@kent.ac.uk}).}
\thanks{Part of this work has been presented in IEEE ICC in London, UK, 2015.}

}

\maketitle\linespread{1.6}

\begin{abstract}
In this paper, we consider the problem of maximizing the weighted sum energy efficiency (WS-EE) for multi-input single-output (MISO) interference channels (ICs) which is well acknowledged as general models of  heterogeneous networks (HetNets), multicell networks, etc. To address this problem, we develop an efficient distributed beamforming algorithm based on a pricing mechanism. Specifically, we carefully introduce a price metric for distributed beamforming design which fortunately allows efficient closed-form solutions to the per-user beam-vector optimization problem. The convergence of the distributed pricing-based beamforming design is theoretically proven. Furthermore, we present an implementation strategy of the proposed distributed algorithm with limited information exchange.  Numerical results show that our algorithm converges much faster than existing algorithms, while yielding comparable, sometimes even better performance in terms of the WS-EE. Finally, by taking the backhaul power consumption into account, it is interesting to show that the proposed algorithm with limited information exchange achieves better WS-EE than the full information exchange based algorithm in some special cases.
\end{abstract}



%
\IEEEpeerreviewmaketitle


\section{Introduction}
In the past few years, various signal processing techniques have been developed to enhance the spectral efficiency (SE) of wireless communication systems \cite{Jinal-2005,Spencer04,Huiling-2009,Huiling-2012,Huilingjsacradio,Jiangzhou,Huilingper,Seung2011,Qingjiang2011}. With these techniques,  the system usually consumes all available power to achieve its maximum SE. Recently, it was reported in \cite{Fettweis-2008} that the power consumption due to  wireless communications is responsible for more than 3\% of the global world $\rm{CO}_2$ emissions and this portion is expected to increase with the soaring demand for high data rate and ever-growing number of mobile devices in the fifth generations (5G) cellular networks \cite{Andrews-2014}. If we do not take urgent measures to conquer this problem, the produced greenhouse gases could bring numerous environmental issues to human beings, such as the sea-level rising, water depletion, the increase of the ocean storm, etc. In addition, most of the mobile terminals are powered by batteries while the development in battery technology is too slow to catch up with the growing demand \cite{Miao-2009}. One promising way to deal with these issues is to optimize the energy efficiency (EE) of future wireless systems, which has received considerable interest in both academia and industry \cite{Yan-2011,Li-2011,Budzisz-2014}.

On the other hand, a key development envisioned in 5G is the evolution of cellular network topology in the shape of the cooperative multicell  \cite{RuiWCNC}, heterogeneous networks \cite{Zhikun2014}, small cells \cite{Andrews-2014}, etc. All these multi-tier networks can be well modeled as an interference channel (IC) where multiple node pairs share the same channel and thus there is mutual interference over the network. Under this scenario, most of the existing studies \cite{Ng2012,Xiaoming2013,Zhikun2014,Quansheng2014,Shiwen-2013,Venturino-2014,zappone2015energy,Zappone2014TSP,weixu} focused on maximizing the global EE (defined as ratio of the sum-SE and the overall power consumption), which is in a fractional form.
The most  popular algorithm to solve these problems is the well-known Dinkelbach method \cite{Dinkelbach-1967,Isheden2012}, which can effectively solve the optimization problem with the fractional objective functions whose numerator and denominator are concave and convex, respectively. Unfortunately, in ICs, the numerator is generally nonconcave due to the presence of multiuser interference and hence the Dinkelbach method does not apply. Various techniques were employed to address this problem, such as the zero-forcing interference elimination technique \cite{Ng2012,Xiaoming2013,Zhikun2014}, orthogonal transmission scheme \cite{Quansheng2014}, alternating optimization technique \cite{Shiwen-2013}, and sequential convex programming \cite{Venturino-2014,zappone2015energy,Zappone2014TSP}.

In practice, however, the global EE may not be suitable to accurately reflect the EE performance of heterogeneous networks (HetNets) since the node pairs cannot share their power and have their own EEs. Moreover, different types of base stations (BSs) are equipped with different types of hardware and thus have different EE requirements. Under this consideration, the metric of weighted sum energy efficient (WS-EE) recently attracts much attentions \cite{Venturino-2014,Venturinotvt,Shiwenjasc,Chenzi-2013,Shiwen-2014}, especially for HetNets.  It is more appealing for this scenario since the weights can be used to control the individual EE on each BS according to their requirements, i.e., giving higher weights for the more energy hungry BSs provides more satisfactory resource-allocation schemes. On the other hand, the global EE is unable to control individual EEs, which is important in HetNets, and it only accounts for the EE of the entire network. Besides, WS-EE provides more tuning abilities to control the EE of the individual BSs. Also, the WS-EE has been widely regarded as the social welfare that was used to measure the performance of the EE games \cite{Meshkati-2005,Lasaulce-2009,Buzzi-2012}.  Recently, there were some studies that modeled WS-EE problem as a pure noncooperative game \cite{nash1951non}, where each node pair maximized its own EE selfishly \cite{Betz-2008,Zappone-2013,Zappone2014TSP,Bacci2014,Miao-2011,bacci2014energy,pan2015totally,zappone2015energy,bo2014}. The EE optimization problem for flat single-input single-output (SISO) ICs was considered in \cite{Betz-2008}. This work was extended in \cite{Zappone-2013} to the case when a relay is present between the transmitters and the receivers. Later on,  \cite{Zappone2014TSP} considered the case where transmitters, receivers and relay are equipped with multiple antennas. In the latter work, all transmitters transmitted only one stream without multiplexing and the relay employed the interference neutralization technique to eliminate the multiuser interference. Study \cite{Bacci2014} took the rate requirements into account and the problem was modeled as a generalized noncooperative game \cite{facchinei2007generalized}. The above-mentioned studies \cite{Betz-2008,Zappone-2013,Zappone2014TSP,Bacci2014} proved the uniqueness (or provided the uniqueness condition) of the (generalized) Nash equilibrium (NE) for the (generalized) noncooperative game by resorting to the standard function framework \cite{Yates1995}, which is applicable when each node pair only needs to optimize one power allocation variable. On the other hand,  both \cite{Miao-2011} and \cite{bacci2014energy} considered the EE optimization problem for frequency-selective SISO ICs without and with rate requirements, respectively. Recently,  we  extended the above studies  to the multiple-input multiple-output (MIMO) scenario with multiplexing \cite{pan2015totally}. In these studies, each node pair has to optimize a power allocation vector or precoding matrix instead of one scalar \cite{Betz-2008,Zappone-2013,Zappone2014TSP,Bacci2014}. Hence, the uniqueness condition  of the NE in these games cannot be derived by using the standard function and the technique based on the contraction mapping \cite{Scutari2008,Scutari-2009} was adopted to obtain the sufficient conditions for the uniqueness of the NE. Most recently, \cite{zappone2015energy} considered the EE optimization problem for both flat and frequency selective SISO ICs, where the minimum-rate constraints were imposed and the signal-to-interference-plus-noise ratio (SINR) was expressed in a more general form.
Although the noncooperative game is an effective tool to devise totally distributed algorithms, the obtained  NE solution  usually suffers  a significant degradation in terms of the overall WS-EE performance since each node pair  maximizes its own EE without considering its negative impact on other users.

There have been some literatures that focus on improving the overall WS-EE performance in ICs \cite{Venturino-2014,Venturinotvt,Shiwenjasc,Chenzi-2013,Shiwen-2014}. In \cite{Venturino-2014}, an iterative algorithm was proposed to directly solve the Karush-Kuhn-Tucker (KKT) conditions of the WS-EE maximization problem for the downlink multicell SISO orthogonal frequency division multiple access (OFDMA) network. However, the proof of the convergence of the iterative algorithm was not provided. In \cite{Shiwenjasc}, the authors considered the WS-EE maximization problem with rate constraints in a heterogeneous multicell network, which was solved by an iterative algorithm based on the weighted minimum mean squared error (WMMSE) \cite{Qingjiang2011}. Although the above literatures significantly improve the overall WS-EE performance, all these algorithms should be implemented in a centralized manner, where  one central control unit (CPU) should exist in the network. All channel information over the network should be conveyed to the CPU, and the CPU calculates all BSs' solutions and send them back to the corresponding BSs. It is well known that the centralized algorithm incurs significant information exchange overhead and high computational complexity, which makes it less scalable to large networks. Therefore, distributed algorithms  appear more desirable and attractive. The authors in \cite{Chenzi-2013} proposed a distributed algorithm for MIMO ICs based on some information exchange among nodes. However, this algorithm was designed based on a critical assumption of symmetric network: the distance from one transmitter to its desired receiver is identical for all node pairs, and each transmitter has the same distance to all its unintended receivers. Also the high-SINR approximation was employed to further simplify the analysis. While for HetNets, node pairs are generated randomly, leading to an asymmetric topology. In addition, these nodes usually operate in intermediate or low SINR regime. Therefore, the algorithm proposed in \cite{Chenzi-2013} is less suitable for HetNets. Recently,  \cite{Shiwen-2014} designed a distributed two-layer EE (TLEE) algorithm to solve the WS-EE optimization problem: the inner layer to update the precoders and decoders; the outer layer to update some parameters. Although it works distributively, the algorithm also incurs much information exchange overhead and high computational complexity due to its two-layer iterative procedure.

In this paper, we aims at jointly optimizing the transmit power and beamforming directions to maximize the WS-EE in multiple-input single-output (MISO) ICs. We presented a distributed algorithm for each node pair to sequentially optimize their beamforming and power based on a pricing mechanism. Enhanced energy efficiency is achieved by our proposed algorithm with reduced information exchange overhead as well as low computational complexity.

The main contributions in this paper are summarized as follows:
\begin{enumerate}
  \item Regarding the nature of ICs, we develop an efficient distributed beam-vector optimization algorithm for maximizing the weighted sum of invidual EE of each link in the ICs. Specifically, a pricing-based WS-EE maximizing problem is carefully designed for each link, and a sequential updating algorithm is presented with its convergence theoretically proven. A reduced signaling overhead strategy with limited information exchange is also presented for practical applications. The performance of the proposed distributed algorithm is  comparable with the centralized algorithm, and better than some existing distributed algorithms.
  \item Depending on the rank of each user's leakage matrix, we divide each user's beamforming optimization problem into two scenarios, for both of which we provide a low-complexity algorithm by exploiting the component decomposition of a beam-vector and applying a linearly combined beam structure. In particular, we factorize each user's beam-vector as the product of the beam direction and power allocation, and optimize them separately. We obtain a near-optimal closed-form expression for the beam directions under both scenarios. For the power allocation problem in the first scenario, we develop an efficient way of finding the globally optimal solution despite its nonconvexity. For the  second scenario, the power allocation problem becomes more complicated. Fortunately, we achieve the globally optimal solution in closed form by solving the KKT conditions. These efforts allow an efficient implementation (computation) of our proposed distributed algorithm for each link.
\end{enumerate}

This paper is organized as follows. In Section \ref{model}, we introduce the system model and the WS-EE maximization problem formulation. Section \ref{ditributed}  presents the distributed adaptive pricing algorithm to solve this problem. In Section \ref{per-link}, we provide a low-complexity algorithm to solve the per-user beam-vector optimization problem. The computational complexity and feedback overhead of the proposed algorithm are compared with existing algorithms in Section \ref{algorthm}. Simulation results are shown in Section \ref{simulation}  and conclusions are  drawn in Section \ref{con}.

\section{System Model and Problem Formulation}\label{model}

We consider an IC with $K$  user pairs, each of which consists of one transmitter equipped with $M$  antennas and one single-antenna receiver. It is assumed that all users are operating over the same frequency band and each receiver suffers from interference from the other  transmitters. Denote ${{\pmb{h}}_{j,k}} \in {\mathbb{C}^{M \times 1}}$ as the flat fading channel vector from the  $j$th transmitter to the $k$th receiver, the elements of which are independently generated with equal variances $\delta_{i,j} $, i.e., $\mathbb{E}\{ {{\pmb{h}}_{j,k}}{\pmb{h}}_{j,k}^{\rm{H}}\}  = \delta_{i,j} {{\pmb{I}}_M}$. It is also assumed that the channels are in block-fading and change sufficiently slowly such that they can be treated as invariant during one transmission period under consideration. The received signal at  the $k$th receiver is given by
\begin{equation}\label{receivedsig}
  {y_k} = {\pmb{h}}_{k,k}^H{{\pmb{w}}_k}{b_k} + \sum\limits_{j \ne k} {{\pmb{h}}_{j,k}^H{{\pmb{w}}_j}{b_j}}  + {z_k},
\end{equation}
where  $b_k$  is the information symbol intended for the $k$th receiver with $\mathbb{E}\{|{b_k}{|^2}\}  = 1$  and $\mathbb{E}\{ {b_k}{b_j}\}  = 0$  for  $k \ne j$, ${{\pmb{w}}_k}$ is user $k$'s beam-vector, and $z_k$ is a zero-mean thermal noise with variance  $\sigma _k^2$.  The SINR of user pair $k$  is given by
\begin{equation}\label{SINR}
  {\eta _k} = \frac{{{{\left| {{\pmb{h}}_{k,k}^{\rm{H}}{{\pmb{w}}_k}} \right|}^2}}}{{\sigma _k^2 + {I_k}}},
\end{equation}
where ${I_k}$  is the interference power received at the  $k$th receiver, given by ${I_k} = \sum\limits_{j \ne k} {{{\left| {{\pmb{h}}_{j,k}^{\rm{H}}{{\pmb{w}}_j}} \right|}^2}}$. Then, the SE (bit/s/Hz) of user $k$  can be written as
\begin{equation}\label{rate}
{r_k}({\pmb{W}}) = {\log _2}\left( {1 + {\eta _k}} \right),
\end{equation}
where ${\pmb{W}} = \left[ {{{\pmb{w}}_1}, \cdots ,{{\pmb{w}}_K}} \right]$  denotes the set of beam-vectors of all transmitters.

To consider the EE design, the total power consumption should be considered for each user pair $k$, which can be modeled as \cite{Shuguang-2004,Saad-2006}
\begin{equation}\label{totalpowercon}
{P_{k,T}} = \rho \left\| {{{\pmb{w}}_k}} \right\|^2 + M{P_{{\rm{ct}},k}}{\rm{ + }}{P_{{\rm{cr}},k}}+P_{\rm{bh}},
\end{equation}
where $\rho  \ge 1$ is a constant accounting for the inefficiency of the power amplifier, ${P_{{\rm{ct}},k}}$  denotes the circuit power consumption for each antenna at  the $k$th transmitter,  ${P_{{\rm{cr}},k}}$ represents the circuit power consumption at the $k$th receiver\footnote{Note that in most existing studies, the circuit power consumption at the receiver is not considered. This is reasonable for cellular systems since user's circuit power consumption is always much lower than the power consumption at the BS side. However, in ICs like  HetNets, the circuit power consumption of receiver is comparable with the transmission energy consumption due to its low transmission power \cite{Shuguang-2004}.}, and $P_{\rm{bh}}$ represents the backhaul power consumption \cite{Saad-2006} which is the sum of the powers consumed by each user to exchange its information with the users within its transmission coverage. For example, denote $d_k$ as the transmission distance for user $k$. The power required for transmission is set by
\begin{equation}\label{perbh}
{P_{k,{\rm{bh}}}} = {\gamma  \cdot {\sigma ^2}}\cdot {{\rm{PL}}_k},
\end{equation}
where $\gamma$ is the target SNR for message exchange, $\sigma ^2$ is the noise variance, and ${\rm{PL}}_k$ is the channel path loss. Then, the EE utility (in bit/Hz/Joule) for user pair $k$  is given by
\begin{equation}\label{normee}
 {U_k}({\pmb{W}})=\frac{{{r_k}({\pmb{W}})}}{{{P_{k,T}}}}.
\end{equation}

Our aim is to develop a distributed algorithm to find the beam-vectors of all users via maximizing the WS-EE of all users, subject to per-user-pair power constraint. Mathematically, this problem can be formulated as
\begin{equation}\label{mainproblem}
\begin{array}{l}
\mathop {\max }\limits_{\pmb{W}} \sum\limits_{k = 1}^K {{\alpha _k}{U_k}({\pmb{W}})} \\
{\rm{s}}{\rm{.t}}{\rm{.   }}\ \left\| {{{\pmb{w}}_k}} \right\|^2 \le {P_k},\forall k,
\end{array}
\end{equation}
where  ${\alpha _k}$'s are weights accounting for priorities of the users, and $P_k$  denotes the maximum transmit power at the $k$th transmitter.

Obviously, problem (\ref{mainproblem}) is nonconvex with respect to (w.r.t.) the set of beam-vectors coupled within the mutual terms. Obtaining the global optimization is known to be difficult even through a centralized algorithm. In the following, we devise a novel distributed algorithm based on the pricing mechanism to deal with this problem.

\section{Distributed Algorithm}\label{ditributed}

In this section, we first introduce a price in the objective function of the per-user beam-vector optimization problem. Based on this, we provide a distributed algorithm where each user pair updates its beam-vector sequentially along with a proof for convergence.

\subsection{Pricing Mechanism}

A proper definition of price is critical for a pricing-based algorithm. In this subsection, we introduce the interference price of user $j$, which represents the marginal decrease in its EE due to a marginal increase in its received interference. It follows
\begin{equation}\label{price}
{\pi _j} \buildrel \Delta \over =  - {\alpha _j}\frac{{\partial {U_j}({\pmb{W}})}}{{\partial {I_j}}} = \frac{1}{{\ln 2}}\frac{{{\alpha _j}{{\left| {{\pmb{h}}_{j,j}^{\rm{H}}{{\pmb{w}}_j}} \right|}^2}}}{{{P_{j,T}}(1 + {\eta _j}){{(\sigma _j^2 + {I_j})}^2}}}.
\end{equation}
When the  $k$th transmitter  transmits data with beam-vector ${{\pmb{w}}_k}$, it generates the interference ${\left| {{\pmb{h}}_{k,j}^{\rm{H}}{{\pmb{w}}_k}} \right|^2}$  to receiver $j$, $j \ne k$. Hence, it is intuitively reasonable to pay a total cost for all other receivers
\begin{equation}\label{cost}
\sum\limits_{j \ne k} {{\pi _j}{{\left| {{\pmb{h}}_{k,j}^{\rm{H}}{{\pmb{w}}_k}} \right|}^2}}  = {\pmb{w}}_k^{\rm{H}}{{\pmb{L}}_k}{{\pmb{w}}_k},
\end{equation}
where  ${{\pmb{L}}_k}$ is called the leakage matrix of the  $k$th transmitter, defined as
\begin{equation}\label{leag}
 {{\pmb{L}}_k} = \sum\limits_{j \ne k} {{\pi _j}{{\pmb{h}}_{k,j}}{\pmb{h}}_{k,j}^{\rm{H}}}.
\end{equation}

In our pricing-based distributed EE design, each transmitter  takes the cost of its generated interference into account. Specifically, given the other transmitters' beam-vectors  ${{\pmb{W}}_{ - k}} = [ {{\pmb{w}}_1}, \cdots ,{{\pmb{w}}_{k - 1}}$, ${{\pmb{w}}_{k + 1}}, \cdots ,{{\pmb{w}}_K} ]$,  the $k$th  transmitter solves the following optimization problem
\begin{equation}\label{peruser}
\begin{array}{l}
\mathop {\max }\limits_{{{\pmb{w}}_k}} \ {{\tilde U}_k}({{\pmb{w}}_k},{{\pmb{W}}_{ - k}}){\rm{ }}\\
{\rm{s}}{\rm{.t}}{\rm{.    }}\ \left\| {{{\pmb{w}}_k}} \right\|^2 \le {P_k},
\end{array}
\end{equation}
where ${\tilde U_k}({{\pmb{w}}_k},{{\pmb{W}}_{ - k}}) \buildrel \Delta \over = {\alpha _k}{U_k}({{\pmb{w}}_k},{{\pmb{W}}_{ - k}}) - {\pmb{w}}_k^H{{\pmb{L}}_k}{{\pmb{w}}_k}$ is the balanced objective by introducing the penalty pricing terms in the desired EE objective of each user pair $k$.

Note that since problem (\ref{peruser}) is still non-convex, obtaining the globally optimal solution is in general a difficult task. The locally optimal solution to problem (\ref{peruser}) can be obtained by some iterative methods in \cite{Dimitri1999}\footnote{It should be emphasized that the iterative methods in \cite{Dimitri1999} can be used to solve the non-convex differentiable problems, including the original WS-EE maximization problem in (\ref{mainproblem}) and the per-link¡¯s optimization problem in (\ref{peruser}). To solve problem (\ref{mainproblem}), the iterative algorithm should be implemented at the central processing unit (CPU) in a centralize manner. The detailed description of the centralized approach is presented in Appendix E, which serves as a performance benchmark for our proposed distributed algorithm. However, the iterative algorithm to solve problem (\ref{peruser}) is implemented at the $k$th transmitter.}. However, the complexity at each transmitter is very high. To reduce the complexity at the transmitters, in Section  \ref{per-link}, we will devise a low-complexity solution with good performance.

\itshape \textbf{Remark 1:}  \upshape The second term in the objective function of (\ref{peruser}) accounts for the cost due to its generated interference, which discourages the  $k$th user pair from maximizing its own EE selfishly. Note that when ${{\pmb{L}}_k} = {\pmb{0}}$, this problem reduces to the non-cooperative game, where each user maximizes its own EE while ignoring its generated interference to  others. The attained NE may not be socially efficient in general. Moreover, it is not clear how to deal with the WS-EE problem when the weights are not equally assigned since each user's optimal solution is independent of the weights through the non-cooperative game theoretical approach.

 \subsection{Distributed Adaptive Pricing Beamforming Algorithm}

 The key idea of the distributed algorithm is that each transmitter has the capability to design its own beam-vector autonomously until convergence. To this end, several updating mechanisms, such as simultaneous updating, sequential updating, or a totally asynchronous updating, can be applied \cite{Bertsekas-1989}.  We present  the sequential version of the algorithm in Algorithm 1, which is referred to as distributed adaptive pricing beamforming (DAPB) algorithm.

 \begin{algorithm}
\caption{Distributed Adaptive Pricing Beamforming (DAPB) Algorithm}
\begin{algorithmic}[1]
\STATE Initialize a feasible  ${{\pmb{W}}^{(0)}} = [{\pmb{w}}_1^{(0)},{\pmb{w}}_2^{(0)}, \cdots ,{\pmb{w}}_K^{(0)}]$,  tolerance $\varepsilon $, iteration number  $n = 1$.  Compute objective value $V_{{\rm{obj}}}^{(0)}{\rm{ = }}\sum\nolimits_{k = 1}^K {{\alpha_k}{U_k}({\pmb{W}}^{(0)})} $.
 \STATE

        \textbf{For}  $k = 1,2, \cdots ,K$,

        \quad Each receiver $j\neq k$ updates its interference price ${\pi _j}$ according to (\ref{price}), and feeds it back to the $k$th

        \quad  transmitter. Having obtained all these interference prices ${\pi _j}\  (j\neq k)$, the $k$th transmitter  solves

         \quad  problem (\ref{peruser}) by using the method developed in Section \ref{per-link}. Denote the solution as ${{\pmb{\tilde w}}_k}$.

          \quad Let ${{\pmb{W}}_{ - k}}= [{\pmb{w}}_1^{(n)},\cdots, {\pmb{w}}_{(k-1)}^{(n)},{\pmb{w}}_{(k+1)}^{(n-1)},\cdots,{\pmb{w}}_{K}^{(n-1)}]$,

          \quad  \textbf{If}  ${\tilde U_k}({{\pmb{\tilde w}}_k},{{\pmb{W}}_{ - k}}) \ge {\tilde U_k}({{\pmb{w}}_k}^{(n-1)},{{\pmb{W}}_{ - k}})$

          \quad\quad The $k$th transmitter updates its beam-vector: ${{\pmb{w}}_k^{(n)}} = {{\pmb{\tilde w}}_k}$;

          \quad  \textbf{else}

       \quad\quad  The  $k$th transmitter keeps its old beam-vector: ${{\pmb{w}}_k^{(n)}} = {{\pmb{w}}_k^{(n-1)}}$;

         \quad \textbf{end}

             \textbf{End}

 \STATE Let ${{\pmb{W}}^{(n)}} = [{\pmb{w}}_1^{(n)},{\pmb{w}}_2^{(n)}, \cdots ,{\pmb{w}}_K^{(n)}]$.  Compute objective  $V_{{\rm{obj}}}^{(n)}{\rm{ = }}\sum\nolimits_{k = 1}^K {{\alpha_k}{U_k}({\pmb{W}}^{(n)})} $. If  ${{\left| {V_{{\rm{obj}}}^{(n)} - V_{{\rm{obj}}}^{(n - 1)}} \right|} \mathord{\left/
 {\vphantom {{\left| {V_{{\rm{obj}}}^{(n)} - V_{{\rm{obj}}}^{(n - 1)}} \right|} {V_{{\rm{obj}}}^{(n - 1)}}}} \right.
 \kern-\nulldelimiterspace} {V_{{\rm{obj}}}^{(n - 1)}}} < \varepsilon $, terminate.  Otherwise, set $n \leftarrow n + 1$  and go to step 2.
\end{algorithmic}
\end{algorithm}

Note that for the sequential updating version of the algorithm, only if  ${\tilde U_k}({{\pmb{\tilde w}}_k},{{\pmb{W}}_{ - k}}) \ge {\tilde U_k}({{\pmb{w}}_k}^{(n-1)},{{\pmb{W}}_{ - k}})$ holds, the $k$th transmitter  will update its beam-vector. Otherwise, it keeps its old beam-vector. This helps guarantee the convergence of the algorithm, as discussed in the following theorem.

 \itshape \textbf{Theorem 1:}  \upshape The sequence of beam-vectors generated by the DAPB algorithm  converges.

 \itshape \textbf{Proof:}  \upshape Please see Appendix A. \hfill $\Box$

\section{Per-transmitter Beam-vector Update}\label{per-link}

In this section, we derive a near-optimal algorithm with low complexity for obtaining the solution to the per-transmitter beam-vector problem in (\ref{peruser}), which is the key step in Algorithm 1.  Based on the rank of the leakage matrix  ${{\pmb{L}}_k}$, we divide problem (\ref{peruser}) into two scenarios: ${\rm{rank(}}{{\pmb{L}}_k}{\rm{)}} = M$ and ${\rm{rank(}}{{\pmb{L}}_k}{\rm{)}} < M$. In the sequel, we deal with the two cases separately.

\subsection{\textbf{Scenario 1:} ${\rm{rank(}}{{\pmb{L}}_k}{\rm{)}} = M$}
Since ${\rm{rank(}}{{\pmb{L}}_k}{\rm{)}} = M$,   ${{\pmb{L}}_k}$ is invertible. Thus, we can introduce an effective  beam-vector ${{\pmb{\bar w}}_k} = {\pmb{L}}_k^{1/2}{{\pmb{w}}_k}$  and define  ${{\pmb{\bar h}}_{k,k}} = {\pmb{L}}_k^{ - 1/2}{{\pmb{h}}_{k,k}}$. Substituting  ${{\pmb{w}}_k} = {\pmb{L}}_k^{ - 1/2}{{\pmb{\bar w}}_k}$ into problem (\ref{peruser}) yields
\begin{equation}\label{changedperuser}
\begin{array}{l}
\mathop {\max }\limits_{{{{\pmb{\bar w}}}_k}} {\rm{ }}\frac{{{\alpha _k}{{\log }_2}\left( {1 + \frac{{{\pmb{\bar w}}_k^{\rm{H}}{{{\pmb{\bar h}}}_{k,k}}{\pmb{\bar h}}_{k,k}^{\rm{H}}{{{\pmb{\bar w}}}_k}}}{{\sigma _k^2 + {I_k}}}} \right)}}{{\rho {\pmb{\bar w}}_k^{\rm{H}}{\pmb{L}}_k^{ - 1}{{{\pmb{\bar w}}}_k} + M{P_{{\rm{ct}},k}} + {P_{{\rm{cr}},k}}+P_{\rm{bh}}}} - {\pmb{\bar w}}_k^{\rm{H}}{{{\pmb{\bar w}}}_k}\\
{\rm{s}}{\rm{.t}}{\rm{.    }}\ {\pmb{\bar w}}_k^{\rm{H}}{\pmb{L}}_k^{ - 1}{{{\pmb{\bar w}}}_k} \le {P_k},\forall k.
\end{array}
\end{equation}
Since this problem is also non-convex, we devise a sub-optimal solution with low complexity.

We decompose ${{\pmb{\bar w}}_k}$  as  ${{\pmb{\bar w}}_k} = \sqrt {{p_k}} {{\pmb{u}}_k}$, where  ${{\pmb{u}}_k}$ can be regarded as its beam direction that is normalized such that its Euclidian norm is unit, and  $p_k$ is the corresponding power. For simplicity, beam direction and power allocation are optimized separately in the following.

\subsubsection{Beam Direction Optimization}

By substituting ${{\pmb{\bar w}}_k} = \sqrt {{p_k}} {{\pmb{u}}_k}$ into (\ref{changedperuser}), we find that beam direction  ${{\pmb{u}}_k}$  only affects the value of the first term of the objective function (FTOF) of problem (\ref{changedperuser}). For simplicity, this beam direction is selected to maximize the numerator of the FTOF of problem (\ref{changedperuser}):  ${{\pmb{u}}_k} = {{{{{\pmb{\bar h}}}_{k,k}}} \mathord{\left/
 {\vphantom {{{{{\pmb{\bar h}}}_{k,k}}} {\left\| {{{{\pmb{\bar h}}}_{k,k}}} \right\|}}} \right.
 \kern-\nulldelimiterspace} {\left\| {{{{\pmb{\bar h}}}_{k,k}}} \right\|}}$.

\subsubsection{Power Allocation Optimization}

When the beam direction ${{\pmb{u}}_k}$  is chosen, the remaining task is to optimize the power component: $p_k$.

For notation simplicity, we define  ${g_{k,k}} \triangleq {{{\pmb{u}}_k^{\rm{H}}{{{\pmb{\bar h}}}_{k,k}}{\pmb{\bar h}}_{k,k}^{\rm{H}}{{\pmb{u}}_k}} \mathord{\left/
 {\vphantom {{{\pmb{u}}_k^{\rm{H}}{{{\pmb{\bar h}}}_{k,k}}{\pmb{\bar h}}_{k,k}^{\rm{H}}{{\pmb{u}}_k}} {(\sigma _k^2 + {I_k})}}} \right.
 \kern-\nulldelimiterspace} {(\sigma _k^2 + {I_k})}}$,  ${A_k} \triangleq {{\rho {\pmb{u}}_k^{\rm{H}}{\pmb{L}}_k^{ - 1}{{\pmb{u}}_k}\ln 2} \mathord{\left/
 {\vphantom {{\rho {\pmb{u}}_k^{\rm{H}}{\pmb{L}}_k^{ - 1}{{\pmb{u}}_k}\ln 2} {{\alpha _k}}}} \right.
 \kern-\nulldelimiterspace} {{\alpha _k}}}$, ${\bar P_k} \triangleq {{{P_k}} \mathord{\left/
 {\vphantom {{{P_k}} {({\pmb{u}}_k^{\rm{H}}{\pmb{L}}_k^{ - 1}{{\pmb{u}}_k})}}} \right.
 \kern-\nulldelimiterspace} {({\pmb{u}}_k^{\rm{H}}{\pmb{L}}_k^{ - 1}{{\pmb{u}}_k})}}$, and ${P_{C,k}} \triangleq {{(M{P_{{\rm{ct}},k}}{\rm{ + }}{P_{{\rm{cr}},k}}+{P_{\rm{bh}}})} \mathord{\left/
 {\vphantom {{(M{P_{{\rm{ct}}}} + {P_{{\rm{cr}}}})} {(\rho {\pmb{u}}_k^{\rm{H}}{\pmb{L}}_k^{ - 1}{{\pmb{u}}_k})}}} \right.
 \kern-\nulldelimiterspace} {(\rho {\pmb{u}}_k^{\rm{H}}{\pmb{L}}_k^{ - 1}{{\pmb{u}}_k})}}$. Problem (\ref{changedperuser}) can be simplified as
	 	
\begin{equation}\label{powerallo}
\begin{array}{l}
\mathop {\max }\limits_{{p_k}} \phi ({p_k}) \buildrel \Delta \over = \frac{{\ln (1{\rm{ + }}{g_{k,k}}{p_k})}}{{{p_k} + {P_{C,k}}}} - {A_k}{p_k}\\
{\rm{s}}{\rm{.t}}{\rm{.    }}\ 0\le {p_k} \le {{\bar P}_k}.
\end{array}
\end{equation}
In the following, we show that problem (\ref{powerallo}) has a unique globally optimal solution, denoted as  ${\mathord{\buildrel{\lower3pt\hbox{$\scriptscriptstyle\frown$}}
\over p} _k}$, given by the following theorem.

 \itshape \textbf{Theorem 2:}  \upshape Denoted $\phi '({p_k})$  as the first derivative of $\phi ({p_k})$  w.r.t. $p_k$. The globally optimal solution ${\mathord{\buildrel{\lower3pt\hbox{$\scriptscriptstyle\frown$}}
\over p} _k}$ of problem (\ref{powerallo}) is characterized  by
\begin{enumerate}
  \item If $\phi '(0) \le 0$, then  ${\mathord{\buildrel{\lower3pt\hbox{$\scriptscriptstyle\frown$}}
\over p} _k} = 0$;
  \item If  $\phi '({\bar P_k}) \ge 0$, then  ${\mathord{\buildrel{\lower3pt\hbox{$\scriptscriptstyle\frown$}}
\over p} _k} = {\bar P_k}$;
  \item If  $\phi '(0) > 0$ and $\phi '({\bar P_k}) < 0$, find  ${\mathord{\buildrel{\lower3pt\hbox{$\scriptscriptstyle\frown$}}
\over p} _k}$ such that  $\phi '({\mathord{\buildrel{\lower3pt\hbox{$\scriptscriptstyle\frown$}}
\over p} _k}) = 0$.
\end{enumerate}

\itshape \textbf{Proof:}  \upshape  Please see Appendix B. \hfill $\Box$

When the optimum solution falls into Case 3), a simple bisection search can be employed to solve it since $\phi '({p_k})$  is monotonically decreasing for  ${\rm{0}} < {p_k} < {\bar P_k}$. It takes at most $\left\lceil {{{\log }_2}({{\bar P}_k}/\varepsilon )} \right\rceil $  iterations to reach tolerance  $\varepsilon $.

\subsection{\textbf{Scenario 2:} ${\rm{rank(}}{{\pmb{L}}_k}{\rm{)}} < M$}

According to (\ref{leag}), condition ${\rm{rank(}}{{\pmb{L}}_k}{\rm{)}} < M$ holds under at least one of the two situations: 1) Sparse networks with $K<M$; 2) The number of positive interference price is smaller than $M$. When ${\rm{rank(}}{{\pmb{L}}_k}{\rm{)}} < M$, the method designed for Scenario 1 is not applicable any more. Hence, a new approach is needed.  For the sake of readability, problem (\ref{peruser}) is written as follows
\begin{equation}\label{rewrittenperuser}
\begin{array}{l}
\mathop {\max }\limits_{{{\pmb{w}}_k}} \frac{{{\alpha _k}{{\log }_2}\left( {1 + \frac{{{{\left| {{\pmb{h}}_{k,k}^{\rm{H}}{{\pmb{w}}_k}} \right|}^2}}}{{\sigma _k^2 + {I_k}}}} \right)}}{{\rho {{\left\| {{{\pmb{w}}_k}} \right\|}^2} + M{P_{{\rm{ct}},k}} + {P_{{\rm{cr}},k}}+{P_{\rm{bh}}}}} - {\pmb{w}}_k^{\rm{H}}{{\pmb{L}}_k}{{\pmb{w}}_k}\\
{\rm{s}}{\rm{.t}}{\rm{.    }}\ \left\| {{{\pmb{w}}_k}} \right\|^2 \le {P_k}.
\end{array}
\end{equation}

In order to develop a low-complexity algorithm to solve problem (\ref{rewrittenperuser}), we first provide a simple form of the beam-vector. Then, the corresponding power allocation is optimized.
\subsubsection{Beam Direction Optimization}
Before providing the simple form of the beam-vector, we first introduce some definitions. Let ${r_k} = {\rm{rank(}}{{\pmb{L}}_k}{\rm{)}} < M$, and apply the eigenvalue decomposition
\begin{equation}\label{SVD}
{{\pmb{L}}_k} = \left[ {{{\pmb{U}}_{k,1}},{{\pmb{U}}_{k,2}}} \right]{{\pmb{\Lambda }}_k}{\left[ {{{\pmb{U}}_{k,1}},{{\pmb{U}}_{k,2}}} \right]^{\rm{H}}},
\end{equation}
where ${{{\pmb{U}}_{k,1}}}$ holds the first ${r_k}$  singular vectors corresponding to the ${r_k}$ positive eigenvalues, and  ${{{\pmb{U}}_{k,2}}}$ holds the last $(M - {r_k})$ singular vectors corresponding to the $(M - {r_k})$ zero-valued eigenvalues. Hence, the orthogonal projection onto the columns of ${{\pmb{L}}_k}$ is ${\Pi _{{{\pmb{L}}_k}}} = {{\pmb{U}}_{k,1}}{\pmb{U}}_{k,1}^{\rm{H}}$, and the orthogonal projection onto the null space of ${{\pmb{L}}_k}$ is $\Pi _{{{\pmb{L}}_k}}^ \bot  = {{\pmb{U}}_{k,2}}{\pmb{U}}_{k,2}^{\rm{H}} = {\pmb{I}} - {{\pmb{U}}_{k,1}}{\pmb{U}}_{k,1}^{\rm{H}}$.
Define two beam directions (with unit norm)
\begin{equation}\label{twodirections}
 {\pmb{w}}_k^{(1)}\triangleq\frac{{{\Pi _{{{\pmb{L}}_k}}}{{\pmb{h}}_{k,k}}}}{{\left\| {{\Pi _{{{\pmb{L}}_k}}}{{\pmb{h}}_{k,k}}} \right\|}},\ {\pmb{w}}_k^{(2)} \triangleq \frac{{\Pi _{{{\pmb{L}}_k}}^ \bot {{\pmb{h}}_{k,k}}}}{{\left\| {\Pi _{{{\pmb{L}}_k}}^ \bot {{\pmb{h}}_{k,k}}} \right\|}},
\end{equation}
which are the projections of channel vector ${{{\pmb{h}}_{k,k}}}$ onto to the space spanned by the columns ${{\pmb{L}}_k}$ and the null space of ${{\pmb{L}}_k}$, respectively. Note that ${\pmb{w}}{_k^{(1){\rm{H}}}}{\pmb{w}}_k^{(2)} = 0$.

Based on the above definitions, we now provide a simple form of the beam-vector as
\begin{equation}\label{simpleform}
{{\pmb{w}}_k} = \sqrt {{p_{k,1}}} {\pmb{w}}_k^{(1)} + \sqrt {{p_{k,2}}} {\pmb{w}}_k^{(2)},
\end{equation}
where ${{p_{k,1}}},{{p_{k,2}}}\geq0$  are the power allocations of beams ${\pmb{w}}_k^{(1)}$ and ${\pmb{w}}_k^{(2)}$, respectively. The choice of the above form of beam-vector takes the following reasons. One arbitrary beam-vector can be divided into two parts: one lies in the space spanned by the columns of ${{\pmb{L}}_k}$; the other one in the null space of ${{\pmb{L}}_k}$. For both parts, the beam directions (with unit-norm) are chosen to maximize the FTOF in (\ref{rewrittenperuser}) or equivalently ${{{\left| {{\pmb{h}}_{k,k}^{\rm{H}}{{\pmb{w}}_k}} \right|}^2}}$ since the beam direction does not affect the denominator of FTOF in (\ref{rewrittenperuser}), which leads to the expression in (\ref{simpleform}). It should be emphasized that the above form of beam-vector is in general not optimal for problem (\ref{rewrittenperuser}), due to the fact that the beam direction in the first part in (\ref{simpleform}) that is chosen to maximize the FTOF of (\ref{rewrittenperuser}) may also maximize the minus part in (\ref{rewrittenperuser}) (i.e., ${\pmb{w}}_k^{\rm{H}}{{\pmb{L}}_k}{{\pmb{w}}_k}$), which may incur a small value of the objective function of (\ref{rewrittenperuser}). However, our simulations show good performance of this heuristic choice of the form of the beam-vector.

\subsubsection{Power Allocation Optimization}
When the beam-vector is chosen as in (\ref{simpleform}), the remaining task is to optimize the power allocations of beams ${\pmb{w}}_k^{(1)}$ and ${\pmb{w}}_k^{(2)}$ (i.e., ${{p_{k,1}}}$ and ${{p_{k,2}}}$).

Using the beam-vector given in (\ref{simpleform}), the desired signal power  at user $k$'s receiver is
\begin{eqnarray}
{\left| {{\pmb{h}}_{k,k}^{\rm{H}}{{\pmb{w}}_k}} \right|^2}&=&{p_{k,1}}{\left| {{\pmb{h}}_{k,k}^{\rm{H}}{\pmb{w}}_k^{(1)}} \right|^2} + {p_{k,2}}{\left| {{\pmb{h}}_{k,k}^{\rm{H}}{\pmb{w}}_k^{(2)}} \right|^2} + 2{\mathop{ Re}\nolimits} \left\{ {\sqrt {{p_{k,1}}{p_{k,2}}} {\pmb{w}}_k^{(2)H}{{\pmb{h}}_{k,k}}{\pmb{h}}_{k,k}^{\rm{H}}{\pmb{w}}_k^{(1)}} \right\}\\
 &\approx& {p_{k,1}}{\left| {{\pmb{h}}_{k,k}^{\rm{H}}{\pmb{w}}_k^{(1)}} \right|^2} + {p_{k,2}}{\left| {{\pmb{h}}_{k,k}^{\rm{H}}{\pmb{w}}_k^{(2)}} \right|^2}\label{approximated}.
\end{eqnarray}
The above approximation is accurate from the average point of view since
\begin{equation}\label{reasonable}
   {\pmb{w}}_k^{(2){\rm{H}}}\mathbb{E}\left\{ {{{\pmb{h}}_{k,k}}{\pmb{h}}_{k,k}^{\rm{H}}} \right\}{\pmb{w}}_k^{(1)} = \delta_{k,k} {\pmb{w}}_k^{(2){\rm{H}}}{\pmb{w}}_k^{(1)} = 0,
\end{equation}
which is also accurate in recent large-scale antenna system.

For notational simplicity, we define ${g_{k,1}} \buildrel \Delta \over = {{{{\left| {{\pmb{h}}_{k,k}^{\rm{H}}{\pmb{w}}_k^{(1)}} \right|}^2}} \mathord{\left/
 {\vphantom {{{{\left| {{\pmb{h}}_{k,k}^{\rm{H}}{\pmb{w}}_k^{(1)}} \right|}^2}} {(\sigma _k^2 + {I_k})}}} \right.
 \kern-\nulldelimiterspace} {(\sigma _k^2 + {I_k})}}$, ${g_{k,2}} \buildrel \Delta \over = {{{{\left| {{\pmb{h}}_{k,k}^{\rm{H}}{\pmb{w}}_k^{(2)}} \right|}^2}} \mathord{\left/
 {\vphantom {{{{\left| {{\pmb{h}}_{k,k}^{\rm{H}}{\pmb{w}}_k^{(2)}} \right|}^2}} {\sigma _k^2}}} \right.
 \kern-\nulldelimiterspace} {(\sigma _k^2 + {I_k})}}$, ${g_{k,3}} \buildrel \Delta \over = {{\rho {\pmb{w}}_k^{(1){\rm{H}}}{{\pmb{L}}_k}{\pmb{w}}_k^{(1)}\ln 2 } \mathord{\left/
 {\vphantom {{\rho {\pmb{w}}_k^{(1){\rm{H}}}{{\pmb{L}}_k}{\pmb{w}}_k^{(1)}} {{\alpha _k}}}} \right.
 \kern-\nulldelimiterspace} {{\alpha _k}}}$ and ${P_{{C,k}}} \triangleq {{\left( {M{P_{{\rm{ct}},k}} + {P_{{\rm{cr}},k}}+{P_{\rm{bh}}}} \right)} \mathord{\left/
 {\vphantom {{\left( {M{P_{{\rm{ct}}}} + {P_{{\rm{cr}}}}} \right)} \rho }} \right.
 \kern-\nulldelimiterspace} \rho }$. By using the approximation in (\ref{approximated}), we consider an alternative problem to (\ref{rewrittenperuser}) as
 \begin{equation}\label{alternative}
\begin{array}{l}
\mathop {\max }\limits_{{p_{k,1}},{p_{k,2}}} \ \frac{{{\ln }(1 + {g_{k,1}}{p_{k,1}} + {g_{k,2}}{p_{k,2}})}}{{{p_{k,1}} + {p_{k,2}} + {P_{C,k}}}} - {g_{k,3}}{p_{k,1}}\\
{\rm{s}}{\rm{.t}}{\rm{.  }}\ {p_{k,1}},{p_{k,2}} \ge 0,{p_{k,1}} + {p_{k,2}} \le {P_k}.
\end{array}
\end{equation}
Obviously, the above problem is a nonconvex problem and more complicated than problem (\ref{powerallo}). Fortunately,  the globally optimal solution to problem (\ref{alternative}) can be obtained by the following theorem.

 \itshape \textbf{Theorem 3:}  \upshape Define three functions
 \begin{equation}\label{function1}
   f_1(x) \triangleq  - (1 + {g_{k,2}}x){\rm{ln}}(1 + {g_{k,2}}x) + {g_{k,2}}(x + {P_{{C,k}}}),
 \end{equation}
 \begin{equation}\label{function2}
 {f_2}(x) \buildrel \Delta \over =  - (1 + {g_{k,1}}x){\rm{ln}}(1 + {g_{k,1}}x) + {g_{k,1}}(x + {P_{{C,k}}}) - {g_{k,3}}{(x + {P_{{C,k}}})^2}(1 + {g_{k,1}}x),
 \end{equation}
 \begin{equation}\label{function3}
 {f_3}(x ) \buildrel \Delta \over = {\rm{ln}}x  + \frac{{{g_{k,2}}{g_{k,3}}}}{{{g_{k,1}} - {g_{k,2}}}}{x ^2} - \ln \left( {\frac{{{g_{k,1}} - {g_{k,2}}}}{{{g_{k,3}}}}} \right).
 \end{equation}
 The globally optimal solution is chosen as the best among all locally optimal solutions. The locally optimal solutions are characterized as follows:
\begin{enumerate}
  \item If $f_1({P_k})<0$ is satisfied, calculate
    \begin{equation}\label{pk2op}
p_{k,2}^\star = \frac{1}{{{g_{k,2}}}}\left[ {{\rm{exp}}\left\{ {\omega \left( {\frac{{{g_{k,2}}{P_{C,k}} - 1}}{e}} \right) + 1} \right\} - 1} \right],
\end{equation}
where function $\omega (\cdot)$ is the Lambert function defined as the inverse function of $f(x) = x{e^x}$\cite{corless1996lambertw}. Compute $\alpha$ as
\begin{equation}\label{afta}
\alpha  = \frac{{\ln (1 + {g_{k,2}}{p_{k,2}^\star})}}{{{{({p_{k,2}^\star} + {P_{{C,k}}})}^2}}} + {g_{k,3}} - \frac{{{g_{k,1}}}}{{({p_{k,2}^\star} + {P_{{C,k}}})(1 + {g_{k,2}}{p_{k,2}^\star})}}.
\end{equation}
If $\alpha\geq0$, one locally optimal solution is given by ${p_{k,1}} = 0$, and ${p_{k,2}} = {p_{k,2}^\star}$.
\item If the following chain of inequalities hold
  \begin{equation}\label{inequallll}
  ({P_k} + {P_{{C,k}}})(1 + {g_{k,1}}{P_k}) > \frac{{{g_{k,1}} - {g_{k,2}}}}{{{g_{k,3}}}} > {P_{{C,k}}},
  \end{equation}
 compute ${f_3}({P_{{C,k}}})$ and ${f_3}({P_k} + {P_{{C,k}}})$. If ${f_3}({P_{{C,k}}}) < 0$ and ${f_3}({P_k} + {P_{{C,k}}}) > 0$, employ the bisection search method to find the solution to  ${f_3}(\theta )=0$. Denote the solution as  $\theta^\star $.  Then, one locally optimal solution is given by
    \begin{equation}\label{locallycase6}
  \left\{ \begin{array}{l}
  {p_{k,1}} = \frac{1}{{{g_{k,1}} - {g_{k,2}}}}\left( {{g_{k,2}}{P_{{C,k}}} + \frac{{{g_{k,1}} - {g_{k,2}}}}{{{g_{k,3}}\theta^\star }} - 1 - {g_{k,2}}\theta^\star } \right)\\
  {p_{k,2}} = \theta^\star  - {P_{{C,k}}} - {p_{k,1}}.
  \end{array} \right.
  \end{equation}
  \item If ${g_{k,1}} > {g_{k,2}}$, compute $\gamma$ as
      \begin{equation}\label{gama}
       \gamma  = \frac{{{g_{k,2}}{g_{k,3}}}}{{{g_{k,1}} - {g_{k,2}}}} - \frac{{\ln ({g_{k,1}} - {g_{k,2}}) - \ln \left( {{g_{k,3}}({P_k} + {P_{{C,k}}})} \right)}}{{{{({P_k} + {P_{{C,k}}})}^2}}}.
      \end{equation}
  If $\gamma\geq0$, we can obtain one locally optimal solution as
   \begin{equation}\label{last}
       \left\{ \begin{array}{l}
        {p_{k,1}} = \frac{1}{{{g_{k,3}}({P_k} + {P_{{C,k}}})}} - \frac{{1 + {g_{k,2}}{P_k}}}{{{g_{k,1}} - {g_{k,2}}}}\\
        {p_{k,2}} = {P_k} - {p_{k,1}}.
       \end{array} \right.
      \end{equation}
  \item  Compute $\gamma$ as
  \begin{equation}\label{gamma}
  \gamma  = \frac{{{g_{k,2}}}}{{({P_k}{\rm{ + }}{P_{{C,k}}})(1{\rm{ + }}{g_{k,2}}{P_k})}} - \frac{{\ln (1 + {g_{k,2}}{P_k})}}{{{{({P_k} + {P_{{C,k}}})}^2}}}.
  \end{equation}
   If $\gamma\geq0$, compute $ \alpha $ as
  \begin{equation}\label{alpha}
  \alpha  = {g_{k,3}} - \frac{{{g_{k,1}} - {g_{k,2}}}}{{({P_k} + {P_{{C,k}}})(1 + {g_{k,2}}{P_k})}}.
  \end{equation}
  If $\alpha\geq0$, one locally optimal solution is ${p_{k,1}} = 0$, ${p_{k,2}} = {P_k}$.
  \item  If two inequalities  ${f_2}(0) > 0,{f_2}({P_k}) < 0$ are satisfied, employ the bisection method to find the solution to equation ${f_2}({p_{k,1}})=0$. The solution is denoted as ${p_{k,1}^\star}$. Compute $\beta$  as
      \begin{equation}\label{beta}
       \beta  = \frac{{\ln (1 + {g_{k,1}}{p_{k,1}^\star})}}{{{{({p_{k,1}^\star} + {P_{{C,k}}})}^2}}} - \frac{{{g_{k,2}}}}{{({p_{k,1}^\star} + {P_{{C,k}}})(1 + {g_{k,1}}{p_{k,1}^\star})}}
      \end{equation}
      If $\beta\geq0$, one locally optimal solution is ${p_{k,1}} = {p_{k,1}^\star},{p_{k,2}} = 0$.
  \item Compute $\gamma$ as
  \begin{equation}\label{PP}
\gamma=- \frac{{\ln (1 + {g_{k,1}}{P_{k}})}}{{{{({P_{k}} + {P_{{C,k}}})}^2}}} + \frac{{{g_{k,1}}}}{{({P_{k}} +\! {P_{{C,k}}})(1\! +\! {g_{k,1}}{P_{k}})}}\! -\! {g_{k,3}}\!
  \end{equation}
  If $\gamma \geq 0$, compute $\beta$ as
   \begin{equation}\label{betaaa}
        \beta  =  \frac{{{g_{k,1}} - {g_{k,2}}}}{{({P_k} + {P_{{C,k}}})(1 + {g_{k,2}}{P_k})}}- {g_{k,3}}.
      \end{equation}
 If $\beta\geq0$, one locally optimal solution is ${p_{k,1}} = {P_k}$, ${p_{k,2}} = 0$.

\end{enumerate}

\itshape \textbf{Proof:}  \upshape Please see Appendix C. \hfill $\Box$

In general, we should check all six cases listed in Theorem 3 and select the best  solution. However,  we can further reduce the complexity by further exploring the properties of these locally optimal solutions.

\itshape \textbf{Lemma 1:}  \upshape If $g_{k,1}\leq g_{k,2}$, the globally optimal solution to problem (\ref{alternative}) is given by
\begin{equation}\label{simplifiedsolution}
  p_{k,1}=0, {p_{k,2}} = \min \left( {\frac{1}{{{g_{k,2}}}}\left[ {\exp \left\{ {\omega \left( {\frac{{{g_{k,2}}{P_C} - 1}}{e}} \right) + 1} \right\}} \right] - 1,{P_k}} \right).
\end{equation}

\itshape \textbf{Proof:}  \upshape Please see Appendix D. \hfill $\Box$

The importance of Lemma 1 is that when $g_{k,1}\leq g_{k,2}$ holds, the globally optimal solution to problem (\ref{alternative}) can be obtained in (\ref{simplifiedsolution}) and there is no need to check all six cases in Theorem 3. This can significantly reduce the search complexity. It should be emphasized that $g_{k,1}\leq g_{k,2}$ is the sufficient condition for the solution given in (\ref{simplifiedsolution}). It can be easily shown that if $g_{k,1}\leq g_{k,2}$ holds, the condition for Case 1 and Case 2 must hold.

\itshape \textbf{Lemma 2:}  \upshape Case 1 and Case 2 cannot happen  simultaneously, neither Case 3 and Case 4.

\itshape \textbf{Proof:}  \upshape  The proof immediately follows  by checking the conditions for these cases, and it is omitted for simplicity. \hfill $\Box$

\section{Algorithm Analysis}\label{algorthm}

\subsection{Complexity Analysis}

For our proposed DAPB algorithm, in each iteration the major complexity lies in the computation of   ${\pmb{L}}_k^{ - 1/2}$
in Scenario 1 or the eigenvalue decomposition of   ${\pmb{L}}_k^{ - 1/2}$ in Scenario 2\footnote{For the power allocation optimization, only some simple scalar mathematical operations are needed. Thus, the complexity of this step can be ignored.}, which involves complexity of  $O({M^3})$ \cite{golub2012matrix}. Hence, the total complexity of the DAPB algorithm is  $O({N_{{\rm{DAPB}}}}K{M^3})$,  where ${N_{{\rm{DAPB}}}}$  denotes the total number of iterations of the DAPB algorithm. For the DBF algorithm in \cite{Chenzi-2013}, the main computational complexity lies in the computation of beam-vector in [41, Eq. (48)], which involves a matrix inversion with complexity  $O({M^3})$. Since in each iteration every link updates its beam-vector, the total complexity of DBF algorithm is  $O({N_{{\rm{DBF}}}}K{M^3})$, where ${N_{{\rm{DBF}}}}$  denotes the total iteration number of DBF algorithm.  As for the TLEE algorithm in \cite{Shiwen-2014}, it consists of two-layer updates: the inner layer for updating beam-vectors  ${\pmb{W}}$, decoders ${\pmb{U}}$  and positive definite matrix  ${\pmb{\Sigma}} $; the outer layer for updating the auxiliary variables  $\pmb{\lambda}$ and $\pmb{\beta}$. Each iteration of the inner layer has a complexity of  $O({K^2}{M^3})$ \cite{Shiwen-2014}. Let ${N_{{\rm{inner}}}}$  and ${N_{{\rm{outer}}}}$  be the number of iterations required in the inner layer and the outer layer, respectively. The TLEE algorithm has a total complexity of  $O({N_{{\rm{inner}}}}{N_{{\rm{outer}}}}{K^2}{M^3})$.
Through simulation tests, we observe that ${N_{{\rm{inner}}}}$  is comparable with ${N_{{\rm{DBF}}}}$,  and ${N_{{\rm{outer}}}}$  is much larger than  ${N_{{\rm{DBF}}}}$. Moreover, the complexity of the TLEE algorithm is quadratic in  $K$, rather than linear in $K$  as the DAPB algorithm and the DBF algorithm. Combing the above two facts and the fact that $K$ is usually large in realistic, the computational complexity of the proposed DAPB algorithm is significantly smaller than that of the TLEE algorithm.   Table \ref{tab1} summarizes the computational complexity comparison.

\begin{table}[!t]
\renewcommand{\arraystretch}{1.1}
\caption{ Computational complexity}
\label{tab1}
\centering
\begin{tabular}{|c|c|c|c|}
\hline
\textbf{Algorithm}  & Full DAPB   & DBF \cite{Chenzi-2013} & TLEE \cite{Shiwen-2014}   \\
\hline
\textbf{Complexity} &  $O({N_{{\rm{DAPB}}}}K{M^3})$ & $O({N_{{\rm{DBF}}}}K{M^3})$ & $O({N_{{\rm{inner}}}}{N_{{\rm{outer}}}}{K^2}{M^3})$ \\
\hline
\end{tabular}
\end{table}

\subsection{Implementation Issues and Overhead Analysis}\label{implementation}

To  implement the DAPB algorithm, each transmitter $k$  needs its own channel vector ${{\pmb{h}}_{k,j}}$, interference plus noise power (IPNP) $I_k+\sigma_k^2$ and the leakage matrix ${{\pmb{L}}_k}$. For its own channel vector, it is assumed the system operates in time-division-duplex (TDD) mode \footnote{In future 5G networks, transmitters are expected to be equipped with large number of antennas \cite{Andrews-2014}. As stated in \cite{Marzetta-2010}, TDD mode is preferable in large-scale MIMO systems due to the fact that its large-dimension channel information is not necessary to be fed back from the receiver. Moreover, this assumption have also been made in some of the existing literatures to design distributed algorithms, e.g., \cite{Shiwen-2014,Chenzi-2013,Qingjiang2011,Dahrouj-2010}. } and the channel vector can be estimated by using the downlink-uplink reciprocity. As for the IPNP at the $k$th receiver, it can be easily measured at the $k$th receiver and then  sent back to the $k$th transmitter. For the leakage matrix ${{\pmb{L}}_k}$, it includes two components according to (\ref{leag}): 1) channel vectors  ${{\pmb{h}}_{k,j}}$, $\forall j \ne k$; 2) interference prices  ${\pi _j}$, $\forall j \ne k$. Channel vectors  ${{\pmb{h}}_{k,j}}$'s can be estimated at the  $k$th  transmitter in a TDD system by exploiting the downlink-uplink reciprocity. For interference prices  ${\pi _j}$ in (\ref{price}), user $j$ needs five quantities:  constant $\alpha_j$, total power consumption $P_{j,T}$, SINR ${\eta _j}$, the variables of its IPNP ${I_j}+\sigma _j^2$, and useful signal power ${\left| {\pmb{h}}_{j,j}^{\rm{H}}{\pmb{w}}_j \right|}^2$. It is assumed that priority $\alpha_j$ is static and has been acquired at the $j$th receiver  before transmission. After the $j$th transmitter  updates its transmission power, it conveys the value of its updated total power consumption $P_{j,T}$ to the $j$th receiver. The $j$th receiver  measures the SINR ${\eta _j}$  and its  IPNP  ${I_j}+\sigma _j^2$. Based on the measurements, the  signal power ${{\left| {{\pmb{h}}_{j,j}^{\rm{H}}{{\pmb{w}}_j}} \right|}^2}$ can be computed as the product of the SINR and the IPNP.
Based on these quantities,  the $j$th receiver can compute its pricing factor ${\pi _j}$ according to (\ref{price}) and broadcast it to the other transmitters.
\begin{figure}
\centering
\includegraphics[width=.45\textwidth]{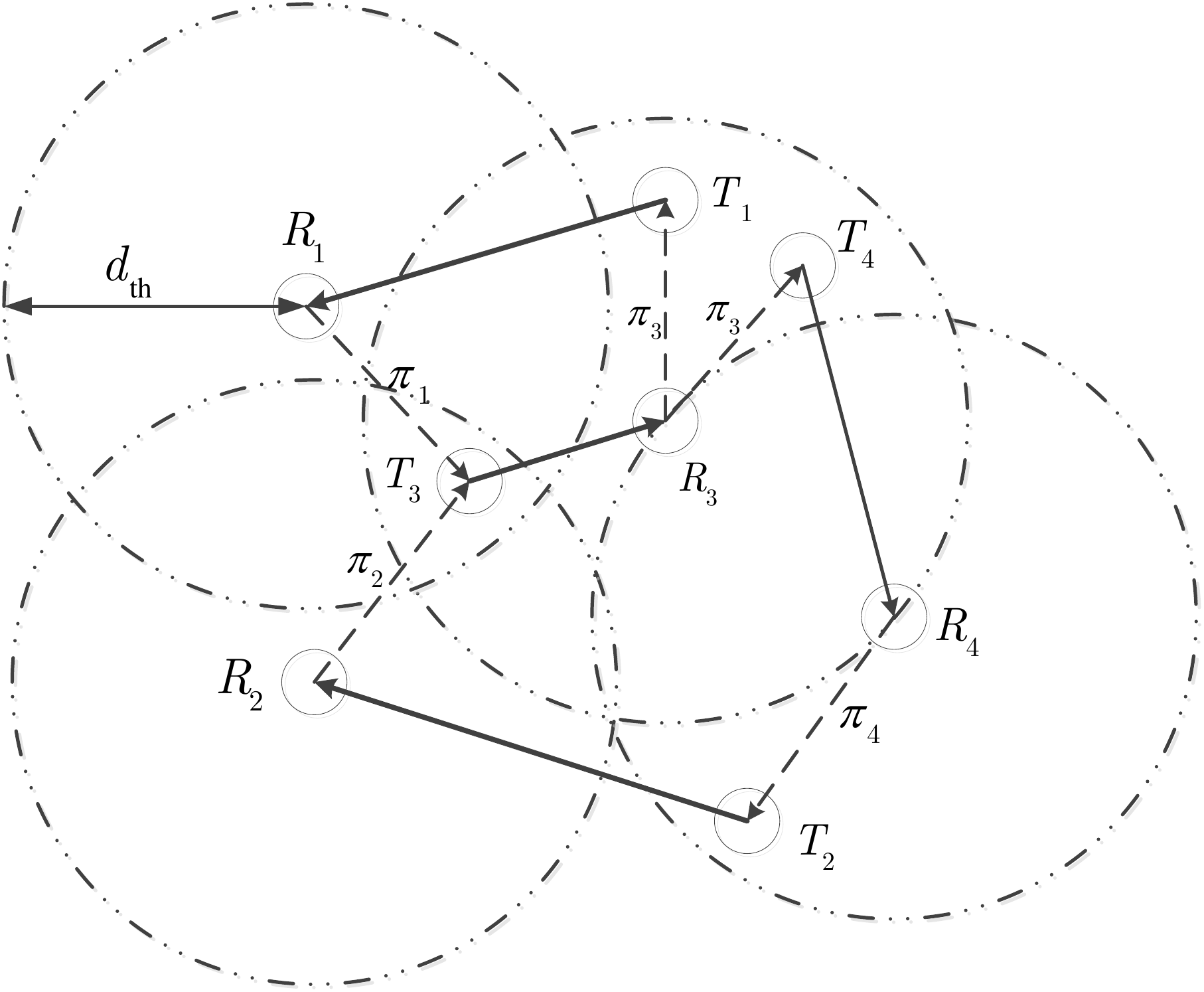}
\caption{An example of implementing the DAPB algorithm with limited information exchange.  `$T_i$' and `$R_i$' represent the $i$th transmitter and receiver, respectively.}
\label{limited-feedback-model}
\end{figure}

According to the above implementation procedure, we are ready to analyze analyze the overhead of the DAPB algorithm. When each user $k$ starts to update its transmit power, it requires its intended receiver to feedback its IPNP, and all other receivers to feedback their interference prices, which incur the feedback of $K$ real scalars. Since all $K$ transmitters are involved in  updating  their transmit power in each iteration, the total overhead of the DAPB algorithm is $N_{\rm{DAPB}}K^2$, where $N_{\rm{DAPB}}$ denotes the number of iterations for the DAPB algorithm to converge. However, we can additionally reduce the overhead of the DAPB algorithm by implementing the DAPB algorithm with limited information exchange, the idea of which is similar to \cite{weijian}. Specifically, if the $k$th transmitter is far away from the $j$th receiver, the channel gain vector ${{\pmb{h}}_{k,j}}$ is very weak, i.e., ${{\pmb{h}}_{k,j}} \approx {\pmb{0}}$. In this case, the interference imposed on the $j$th receiver from the $k$th transmitter becomes negligible. Under this consideration, there is no need for the $j$th receiver to feedback the interference price $\pi_j$ to the $k$th transmitter. Hence, we assume that each receiver only broadcasts the interference price to its nearby transmitters within a specified distance $d_{\rm{th}}$. Fig.~\ref{limited-feedback-model} illustrates an example of four user pairs. The dashed arrows represent the channel to broadcast the price information, while the solid arrows are the signaling channels.
 From this example, we see that $R_1$ only sends the price $\pi_1$ to its close terminal $T_3$ within the distance of $d_{\rm{th}}$, instead of all terminals. Similar observations can be seen for the price information exchange at $R_2$ and $R_3$. This limited information exchange method can significantly reduce the backhaul power consumption according to  (\ref{perbh})  and the simulation section will evaluate the performance of this method. This method is referred to as the limited DAPB algorithm. To evaluate the overhead of this algorithm, we first introduce some notations. Denote the set of all receivers that need to feedback information (including pricing information and its own IPNP) to transmitter $k$ as ${\cal T}_k$.  Denote the cardinality of  ${\cal T}_k$ as $N_k$, i.e., $N_k=|{\cal T}_k|$. In general, $N_k$ is much smaller than $K$. The total overhead of the limited DAPB algorithm is reduced to $N_{\rm{lim-DAPB}}\sum\nolimits_{k = 1}^K N_k$, where $N_{\rm{lim-DAPB}}$ is the number of iterations for its convergence.

For comparison, we here characterize the overhead by the centralized algorithm as detailed in Appendix E. The centralized algorithm generally requires a central processing unit (CPU) collecting all channel vectors within the network for centralized optimization. This incurs the feedback of $2K^2M$ real scalars corresponding to all the channel information. The CPU computes all transmitters' beam-vectors and conveys them back to the corresponding transmitters, which incurs the further beamforming information exchange of $2KM$ real scalars. Hence, the total overhead of the centralized algorithm is $2K^2M+2KM$. Note that the overhead of the centralized algorithm significantly depends on the number of antennas in contrast to the proposed (limited) DAPB algorithm, which is not related to the number of antennas. In future 5G system, some transmitters are expected to be equipped with large-scale antenna array \cite{Andrews-2014} and it could cause a prohibitively high overhead burden.

Concerning the noncooperative algorithm,  no extra information exchange is necessary among different user pairs. The only information that needs to be exchanged is their own IPNP, that is, the $k$th transmitter only requires its intended receiver to feed back its IPNP. Given $K$ transmitters in the network, the total overhead of the noncooperative algorithm is $N_{\rm{Nonco}}K$, where $N_{\rm{Nonco}}$ denotes the number of iterations for its convergence.

Now we also characterize the overhead of several modified algorithms proposed in recent years. For DBF algorithm, each user's receiver needs to feed back its weighted receiver filter \cite{Chenzi-2013}, which is a complex scalar. Hence, the DBF algorithm requires a total number of  $2{N_{{\rm{DBF}}}}K$ real numbers, where $N_{{\rm{DBF}}}$ is the total number of iterations for the DBF algorithm to converge. For the TLEE algorithm, we first consider the inner layer (step 2- step 5 in Algorithm 1 of \cite{Shiwen-2014}), then the outer layer (step 6 in Algorithm 1 of \cite{Shiwen-2014}). In the inner layer, each receiver $j$  feeds back the complex receiver filter  ${u_j}$, the updated real number ${\sum _j}$  and the signal covariance scalar ${J_j}$  to its transmitter. Moreover, each user needs to broadcast one real number to the other transmitters in order to calculate the objective value  $\rho $ in step 5 of Algorithm 1 of \cite{Shiwen-2014}. Hence, each inner iteration needs $5K$ real numbers for exchange. For the outer layer, according to [\cite{Shiwen-2014}, Eqs. (34a)-(34c)], each user needs to broadcast about five real numbers to calculate the summation in [\cite{Shiwen-2014}, Eq. (37)]. Combing the above two facts, the TLEE algorithm requires a total of $5K{N_{{\rm{out}}}} + 5K{N_{{\rm{in}}}}{N_{{\rm{out}}}}$  real numbers, where ${N_{{\rm{in}}}}$  and ${N_{{\rm{out}}}}$  denote the number of iterations required in the inner layer and the outer layer, respectively.

Table \ref{tab2} summarizes the above overhead comparison, where `Full DAPB' refers to the DAPB algorithm with full information exchange.

\begin{table}[!t]
\renewcommand{\arraystretch}{1.1}
\caption{ Overhead Comparison (Number of real scalars)}
\label{tab2}
\centering
\begin{tabular}{|c|c|c|c|c|c|c|}
\hline
\textbf{Algorithm}  & Full DAPB  & limited DAPB & Centralized &  Noncoop  & DBF & TLEE \\
\hline
\textbf{Overhead} & ${N_{{\rm{DAPB}}}}K^2$ & $N_{\rm{lim-DAPB}}\sum\nolimits_{k = 1}^K N_k$ & $2K^2M+2KM$ & $N_{\rm{Nonco}}K$ & $2{N_{{\rm{DBF}}}}K$ & $5KN_{\rm{out}} (N_{\rm{in}}+1)$   \\
\hline
\end{tabular}
\end{table}

\section{Simulation Results}\label{simulation}

In this section, we present simulation results to evaluate the performance of the proposed DAPB algorithm.  We consider a network contained in a square of length $d_{\rm{len}}=350$ meters, within which all users are randomly dropped. The distance between the transmitter and its designated receiver is uniformly distributed in the range of $[30{\rm{m}},60{\rm{m}}]$.  Also, each transmitter has at least  $30{\rm{m}}$ distance from its unintended receiver. All channels are modeled by the product of  path-loss and independent Rayleigh fading with complex normal distribution ${\cal C}{\cal N}(0,1)$. The path loss in  decibels is modeled as $38.46+35{\rm{log}}_{10}(d)$ \cite{Assumptions-2009}.
 Each channel realization is obtained by generating an random set of user positions as well as fading channel realizations. All  simulation results are obtained by averaging over 1000 channel realizations. Unless stated otherwise,  we assume equal maximum transmit power (i.e., ${P_k} = {P_{\max }},\forall k$) and equal weights (i.e., $\alpha_k=1, \forall k$) for all users.  To account for the features of the heterogeneous networks, we assume the circuit power per antenna at TX (i.e., $P_{{\rm{ct}},k}, \forall k$) and RX (i.e., ${P_{{\rm{cr}},k}}, \forall k$) are randomly generated within [50mW, 200mW] and [200mW, 400mW], respectively. The other main system parameters are given in Table \ref{tab3}.  Normalized channel matched beamformers are employed as initial beam directions and  power allocations for all users are randomly initialized. Although different initial beam-vectors may lead to different locally optimal solutions due to the non-convexity of the original problem in (\ref{mainproblem}), the above simple initialization method performs very well as shown in the following results.


\begin{table}[!t]
\renewcommand{\arraystretch}{1.1}
\caption{Main simulation parameters}
\label{tab3}
\centering
\begin{tabular}{c|c}
\hline\hline
\textbf{Parameters}  & \textbf{Value}  \\
 \hline
 Tolerance $\varepsilon $  & $10^{-3}$     \\
 \hline
Number of antennas  $M$  & 4    \\
\hline
Noise power spectral density  & -174 dBm/Hz \cite{Shuguang-2004} \\
\hline
Channel Bandwidth  $B$ &  20 MHz \\
\hline
Power amplifier efficiency & ${1 \mathord{\left/ {\vphantom {1 \rho }} \right.\kern-\nulldelimiterspace} \rho } = 0.35$ \cite{Shuguang-2004}\\
\hline
Target SNR for message exchange $\gamma$ & 4 dB \cite{Saad-2006} \\

\hline\hline
\end{tabular}
\end{table}

\subsection{Properties of the DAPB algorithm}

\subsubsection{Convergence Behavior}

\begin{figure}
\centering
\includegraphics[width=4in]{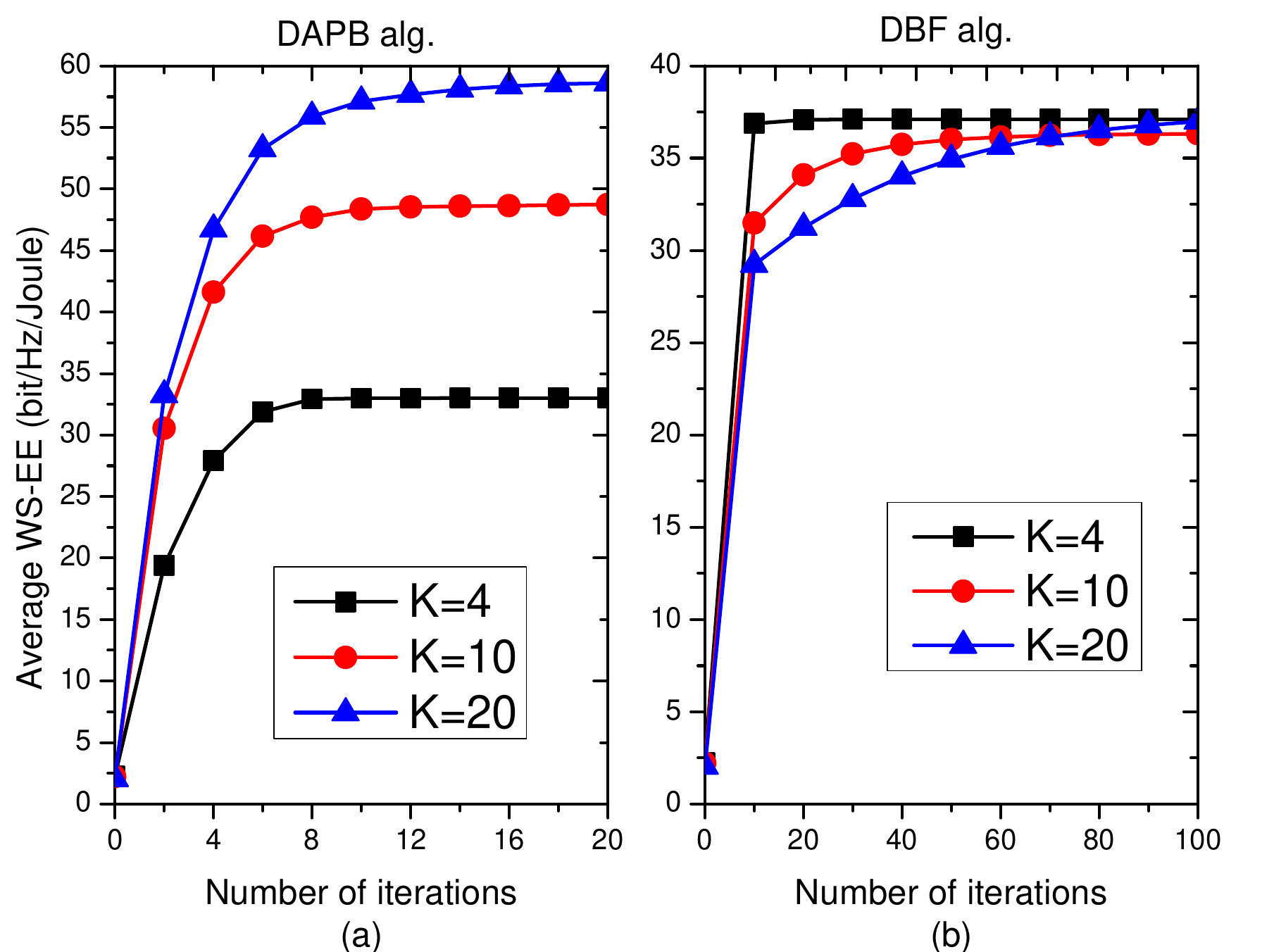}
\caption{Convergence behaviors of the DAPB algorithm (a) and DBF  algorithm (b) under different number of users.}
\label{convergence-DAPBandDBF}
\end{figure}

\begin{figure}
\centering
\includegraphics[width=4in]{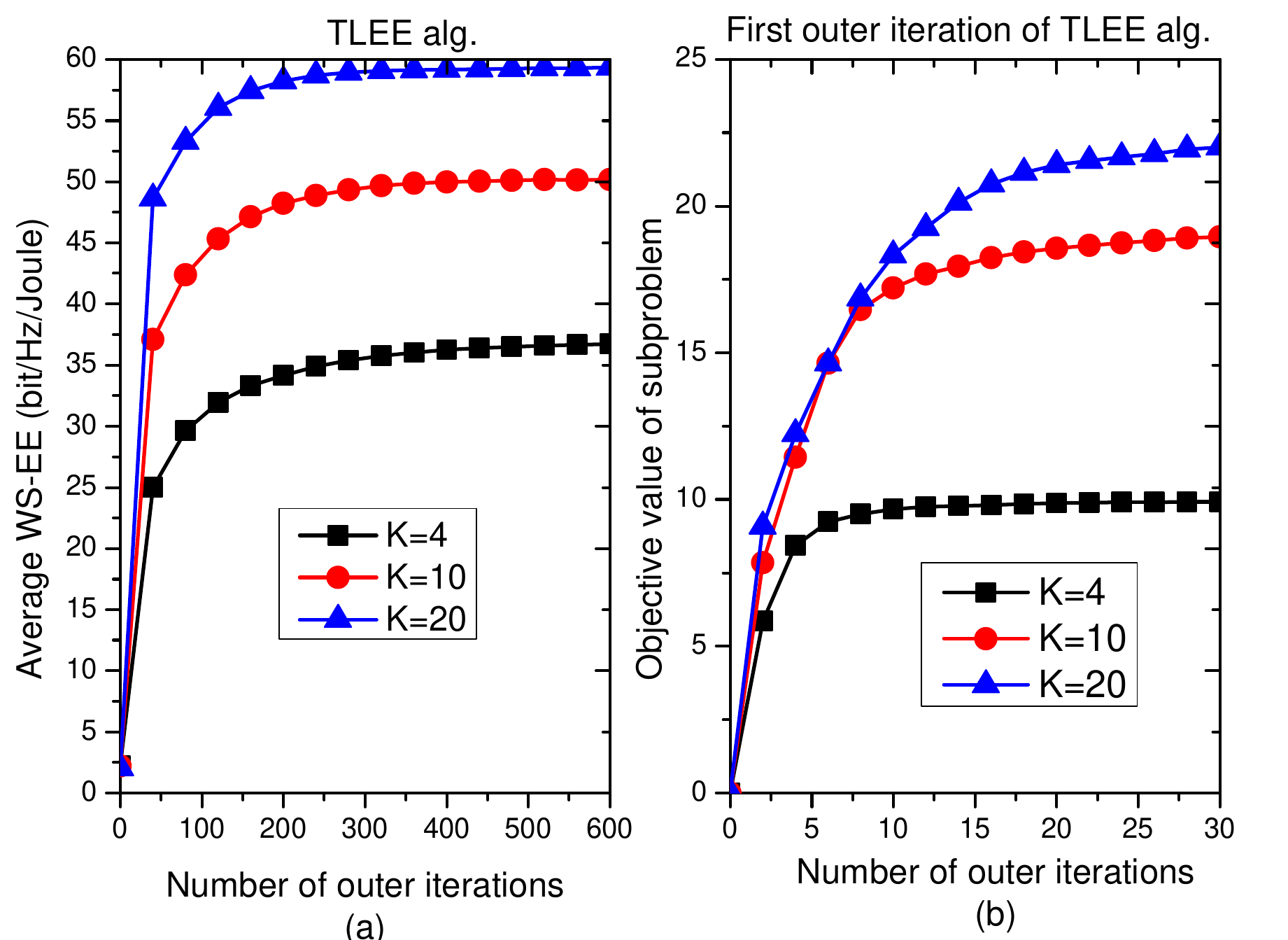}
\caption{(a): Convergence behavior of the TLEE algorithm; (b): Convergence behavior of the inner updates in the first outer iteration of the TLEE algorithm. }
\label{convergence-TLEE}
\end{figure}

We first compare the convergence speed of the proposed DAPB algorithm with the existing distributed algorithms: DBF algorithm \cite{Chenzi-2013} and TLEE algorithm \cite{Shiwen-2014}.
Fig.~\ref{convergence-DAPBandDBF} and Fig.~\ref{convergence-TLEE} illustrate the convergence behaviors of different algorithms for different numbers of users with $P_{\rm{max}}=33 \ {\rm{dBm}}$. Specifically,  Figs.~\ref{convergence-DAPBandDBF} (a) and (b) correspond to the DAPB algorithm and the DBF algorithm, while Figs.~\ref{convergence-TLEE} (a) and (b) correspond to the outer and inner layers of the TLEE algorithm. It can be seen from Fig.~\ref{convergence-DAPBandDBF} (a) that the WS-EE monotonically increases when iterating  the DAPB algorithm and it converges rapidly (within 20 iterations for all considered configurations). On the other hand, from  Fig.~\ref{convergence-DAPBandDBF} (b), it can be seen that the convergence speed of the DBF algorithm is significantly affected by the number of users: A larger $K$ corresponds to slower convergence. In the case of $K=20$, the DBF algorithm requires more than 100 iterations to converge. Moreover, from the right panel of Fig.~\ref{convergence-TLEE}, we find that the TLEE algorithm has the worst convergence performance and needs more than 400 iterations to converge for all considered configurations. In addition, some inner iterations are required in each outer iteration as seen in  Fig.~\ref{convergence-TLEE} (b). Hence, our proposed DAPB algorithm has the least number of iterations and thus the lowest computational complexity according to Table \ref{tab1}.

\subsubsection{Impacts of maximum power on convergence speed}
\begin{figure}
\centering
\includegraphics[width=4in]{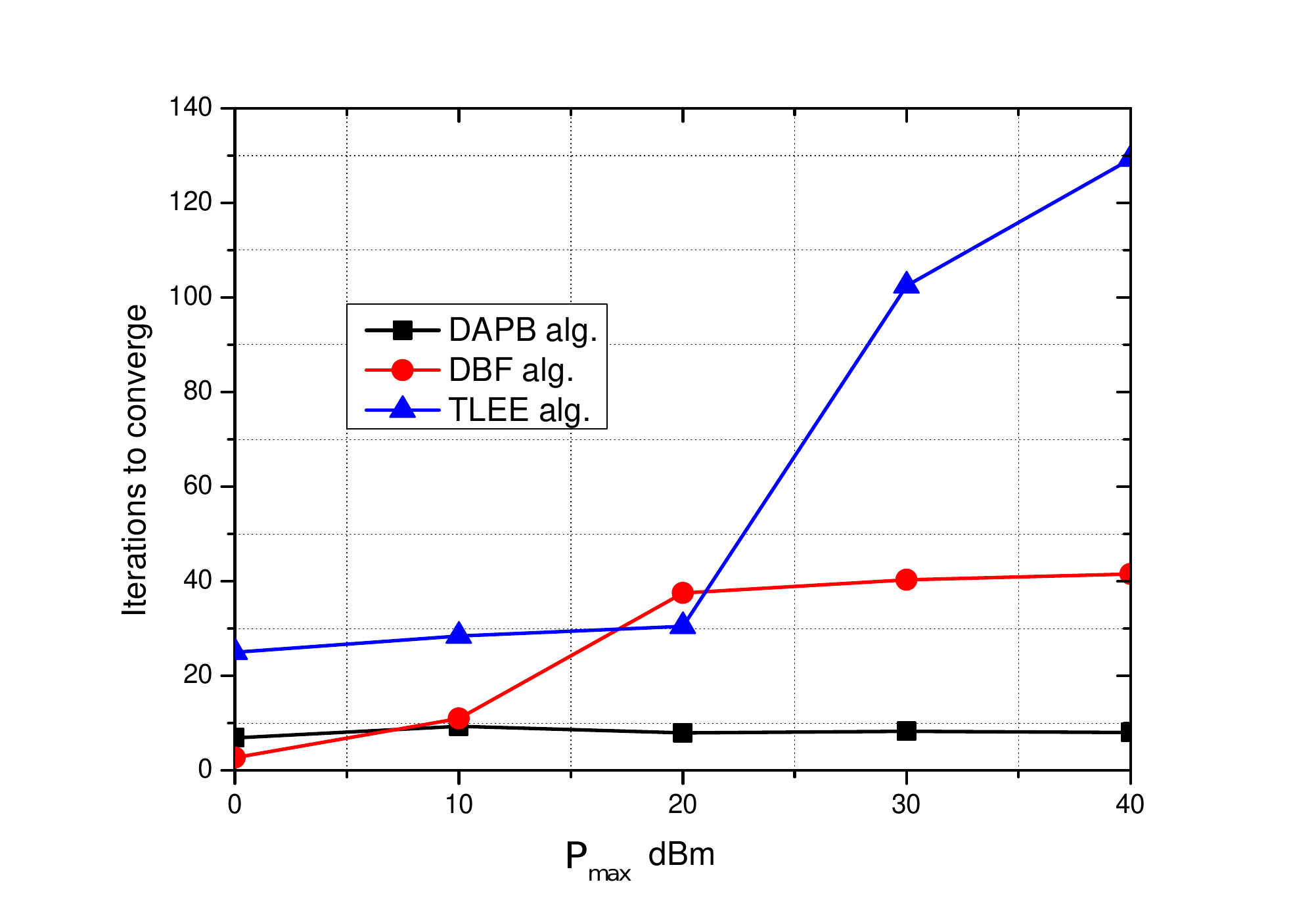}
\caption{Convergence speed versus $P_{\rm{max}}$ under various algorithms. The number of users is $K=10$.}
\label{Convergence-speed}
\end{figure}
In Fig.~\ref{Convergence-speed}, we investigate the effect of the maximum transmit power on the convergence speed of various distributed algorithms. The figure shows that the number of iterations required for the TLEE algorithm increases dramatically with the increase of the maximum transmit power, while the number of iterations required for our proposed DAPB algorithm stays roughly fixed (within 10 iterations) for all  transmit power constraints.

\subsubsection{Comparison of the computational time }
\begin{table}[!t]
\renewcommand{\arraystretch}{1.1}
\caption{Average computational time (in seconds) for various methods.}
\label{tab4}
\centering
\begin{tabular}{|c|c|c|c|c|c}
\hline
  & DAPB  &  DBF \cite{Chenzi-2013}  &  TLEE \cite{Shiwen-2014} &  Centralized \\
\hline
$K=4$ &  0.019  &  0.046  &    1.676 &    2.12    \\
 \hline
 $K=20$  & 0.170 &  7.456 &   5.696 &   72.4  \\
\hline
\end{tabular}
\end{table}

In Table \ref{tab4},  we report  the computational time of various distributed algorithms when applied to the same scenarios as those in Fig.~\ref{convergence-DAPBandDBF} and Fig.~\ref{convergence-TLEE}. For comparison, we also report the computational time for an alternative centralized algorithm  given in Appendix E. Our simulation configuration is E5-1650 CPU with 3.5GHz. To ensure that the centralized algorithm has a good performance, its tolerance should be set as $10^{-5}$. As expected, the centralized algorithm requires significantly more time than the distributed algorithms, especially in dense networks. Hence, the centralized algorithm has the highest computational complexity. It is seen from the table that the DBF and TLEE algorithms require much more time than our proposed DAPB  algorithm, and they become prohibitively high in the case of $K=20$. Also, the computational time of the DBF algorithm is significantly affected by the number of users. For example, when the number of users increases from $K=4$ to $K=20$, the computational time of the DBF algorithm increases up to over 100 times, in contrast to 10 times for the DAPB algorithm and 6 times for the TLEE algorithm. Note that even for the smaller number of users with $K=4$, the TLEE algorithm also requires much computational time (i.e., 2.6s), and is even comparable with the centralized algorithm.

\subsubsection{Impacts of the weights}

\begin{figure}
\centering
\includegraphics[width=4in]{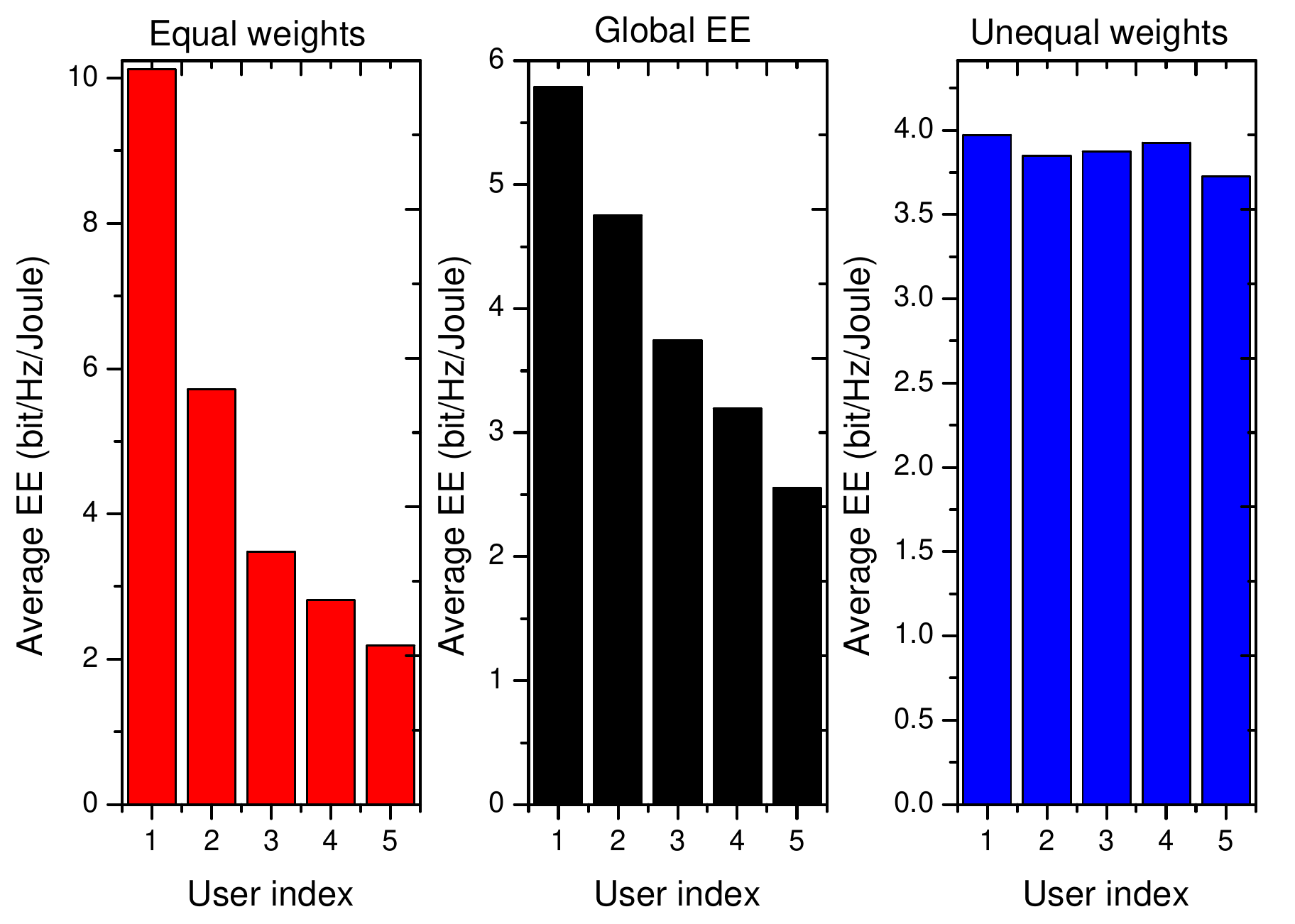}
\caption{Individual EE distribution under different schemes. (Left plot) Equal weights: $\alpha_k=0.2$, for $k=1,\cdots,5$; (Middle plot) Global EE maximization method \cite{Shiwen-2013}; (Right plot) Unequal weights: $\alpha_1=0.02, \alpha_2=0.03, \alpha_3=0.07, \alpha_4=0.25, \alpha_5=0.63$.}
\label{EE-weights}
\end{figure}

As mentioned, the weights can be used to give priority to the specific user pairs, which is appealing in heterogeneous networks. To give more insights, we consider a network composed of five users pairs with $P_{{\rm{ct}},1}=50{\rm{mW}}, P_{{\rm{ct}},2}=90{\rm{mW}}, P_{{\rm{ct}},3}=130{\rm{mW}}, P_{{\rm{ct}},4}=160{\rm{mW}}, P_{{\rm{ct}},5}=200{\rm{mW}}$ and $P_{{\rm{cr}},1}=200{\rm{mW}}, P_{{\rm{cr}},2}=250{\rm{mW}}, P_{{\rm{cr}},3}=300{\rm{mW}}, P_{{\rm{cr}},4}=350{\rm{mW}}, P_{{\rm{cr}},5}=400{\rm{mW}}$. We test two sets of weights: 1) $\alpha_k=0.2,\forall k$; 2) $\alpha_1=0.02, \alpha_2=0.03, \alpha_3=0.07, \alpha_4=0.25, \alpha_5=0.63$. Moreover, we also consider the performance of the global EE maximization method provided in \cite{Shiwen-2013}.
In Fig.~\ref{EE-weights}, we show the users' EE distribution under different schemes. As expected, for the cases of equal weights and global EE maximization, user 1 achieves the highest EE since it has the lowest  circuit power consumption, whereas the lowest EE is achieved for user 5 since it has the highest circuit power consumption. In these two cases, the obtained resource-allocation solutions present a significant spread of the individual EE across the users. On the other hand, for the case of unequal weights, a more balanced EE distribution can be achieved by assigning higher weights to the users with higher circuit power consumption and lower weights to the ones with lower circuit power consumption. Hence, compared with the global EE maximization method, WS-EE is more flexible to control individual EEs and fairness among the users can be guaranteed by appropriately tuning the set of weights.

 \subsection{Performance Comparison}

 We next compare the performance of the proposed DAPB algorithm with the DBF algorithm \cite{Chenzi-2013}, the TLEE algorithm \cite{Shiwen-2014}, the centralized algorithm based on gradient projection method given in Appendix E, as well as the global EE maximization method in \cite{Shiwen-2013}. For fairness, we assume that the backhaul power consumptions of the above five algorithms are identical. Also, the performance of  the non-cooperative algorithm is considered, where each transmitter maximizes its own EE selfishly. For comparison, the performance of the limited DAPB algorithm proposed in Subsection \ref{implementation} is also illustrated, for which each receiver only broadcasts the interference price to its nearby transmitters within a specified distance $d_{\rm{th}}$. In general, $d_{\rm{th}}$ is much smaller than the length of this region. Hence, this method incurs much less backhaul power consumption compared with other algorithms except the noncooperative algorithm, where each receiver is only necessary to feedback information its serving transmitter.

\begin{figure}
\centering
\includegraphics[width=4in]{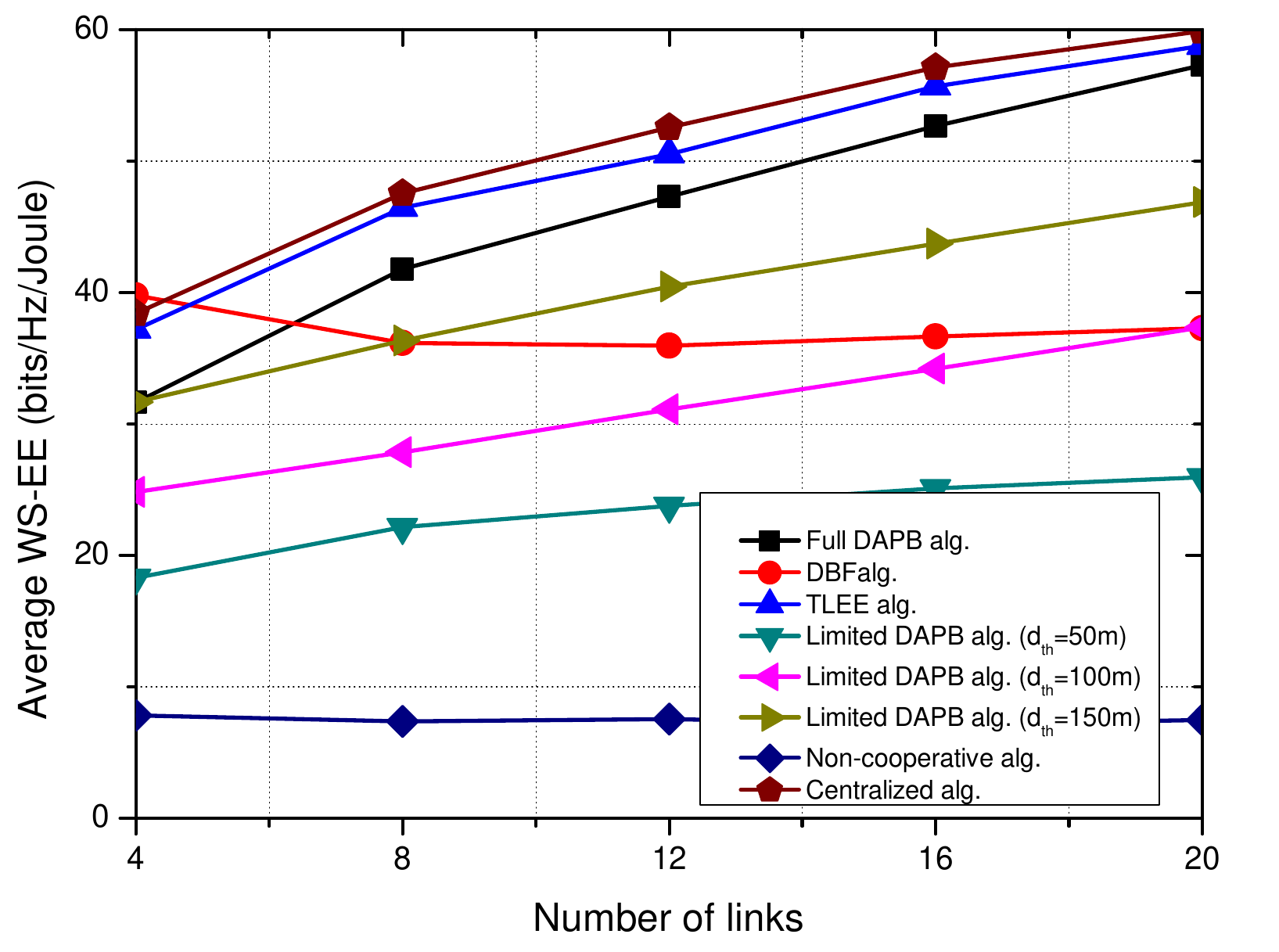}
\caption{Average WS-EE versus the number of links under various algorithms.}
\label{EE-K}
\end{figure}

 Fig.~\ref{EE-K} shows the average WS-EE under various algorithms versus the number of links. The full DAPB algorithm refers to the DAPB algorithm with full information exchange. By using the proposed pricing mechanism to regulate the interference, the full DAPB algorithm significantly outperforms the non-cooperative algorithm in the sense of offering higher WS-EE. The performance gain improves with the increase of the number of links. This is due to the fact that as the number of links increases, the users suffer from high interference, which can be efficiently mitigated by the pricing mechanism in the full DAPB algorithm. We also notice that the performance of the full DAPB algorithm is better than that of the DBF algorithm, and the performance gain is significant when the number of links is large. The reason is that the DBF algorithm is primarily designed for the symmetrical system and may not be suitable for the asymmetrical network considered here. Also, increasing the number of links will make the network more asymmetrical. It is observed that the full DAPB algorithm achieves a similar WS-EE performance as the TLEE algorithm.  Numerical results also show that the centralized algorithm provides marginal performance gains over the DAPB algorithm and the TLEE algorithm. However, it incurs the highest computational complexity as seen from Table \ref{tab3}. As expected, the performance of the limited DAPB algorithm is worse than the full DAPB algorithm due to the interfering users farther away than the transmission distance.

\begin{figure}
\centering
\includegraphics[width=4in]{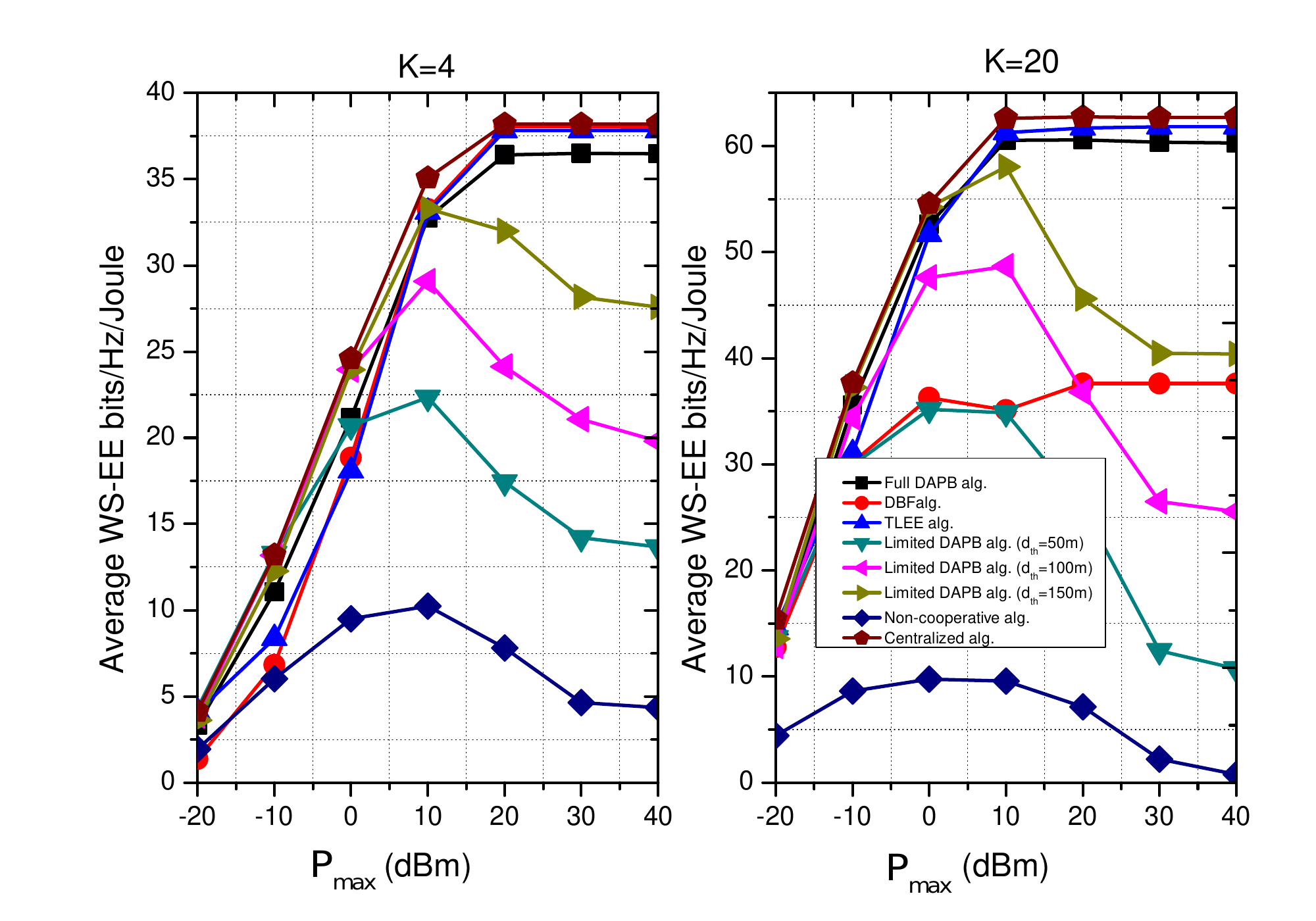}
\caption{Average WS-EE  versus the maximum transmit power under various algorithms for (a) $K=4$  and (b) $K=20$.}
\label{EE-Pmax}
\end{figure}

In Figs.~\ref{EE-Pmax} (a) and (b), we illustrate the WS-EE versus the maximum transmit power under various algorithms for $K=4$ and $20$, respectively. It is observed that the WS-EE of the full DAPB algorithm increases with the transmit power constraint. While in the high transmit power regime, its WS-EE becomes saturated. This is because in this regime the full DAPB algorithm is unwilling to increase its transmit power to further increase SE in order to maximize the WS-EE. On the other hand, for the limited DAPB and noncooperative algorithms, the WS-EE first increases with the transmit power  and then decreases in the high transmit power regime. The reason is as follows: In the low transmit power regime, network interference so weak that it is negligible, while in the high transmit power regime network interference becomes too strong to be neglected.  It is interesting to find that in the low transmit power regime such as -20 $\sim$  0 dBm for $K=4$ and -20 $\sim$ -10 dBm for $K=20$, the limited DAPB algorithm is even better than the full DAPB algorithm in terms of the WS-EE performance. This is mainly due to the fact that in the low transmit power regime, the backhaul power consumption will become a dominant factor for limiting the system EE performance and the backhaul power consumption of the limited DAPB algorithm is smaller than the full DAPB algorithm.  This result suggests that in some scenarios our proposed limited DAPB algorithm is a better option in terms of the system EE performance. Similarly to that observed  in Fig.~\ref{EE-K}, the DBF algorithm performs well in the sparse network with $K=4$, while appears worse in the dense network with $K=20$.

\begin{figure}
\centering
\includegraphics[width=4in]{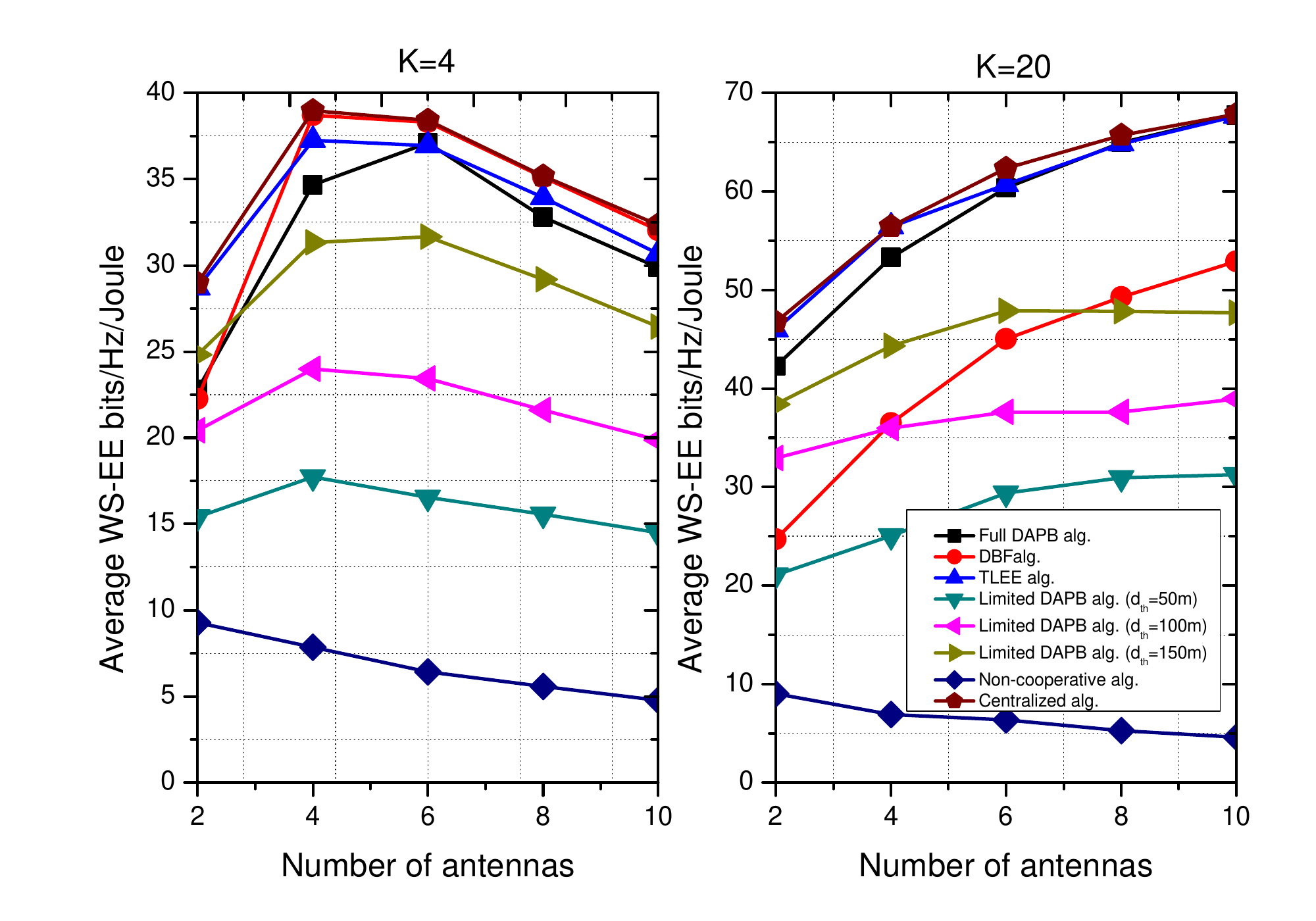}
\caption{Average WS-EE versus the number of antennas under various algorithms  for (a) $K=4$  and (b) $K=20$.}
\label{EE-M}
\end{figure}

Figs.~\ref{EE-M} (a) and (b)  depict the average WS-EE versus the number of antennas under various algorithms for $K=4$ and $20$, respectively. In the sparse network with $K=4$, WS-EE achieved by all the algorithms (except the non-cooperative algorithm) first increase with the number of antennas and then decrease when the number of antennas is larger. On the other hand, when $K=20$, the average WS-EE achieved by all algorithms always increase for all considered numbers of antennas. The reason is that when the number of links is small, the interference is not so important. Hence, only a limited number of antennas is enough to mitigate the interference. Additional antennas provide marginal gains in the SE but incurs significant circuit power consumption, leading to the decrease of the WS-EE. In the dense network, however, interference becomes the dominant factor for limiting the system performance. In this case, more degrees of freedom (i.e., antennas) are helpful to enhance the signal quality of each user while causing minimum amount of interference to the other users, leading to significant SE enhancement. This SE improvement overwhelms the increased circuit power consumption, and results in the increase of the WS-EE.

\begin{figure}
\centering
\includegraphics[width=4in]{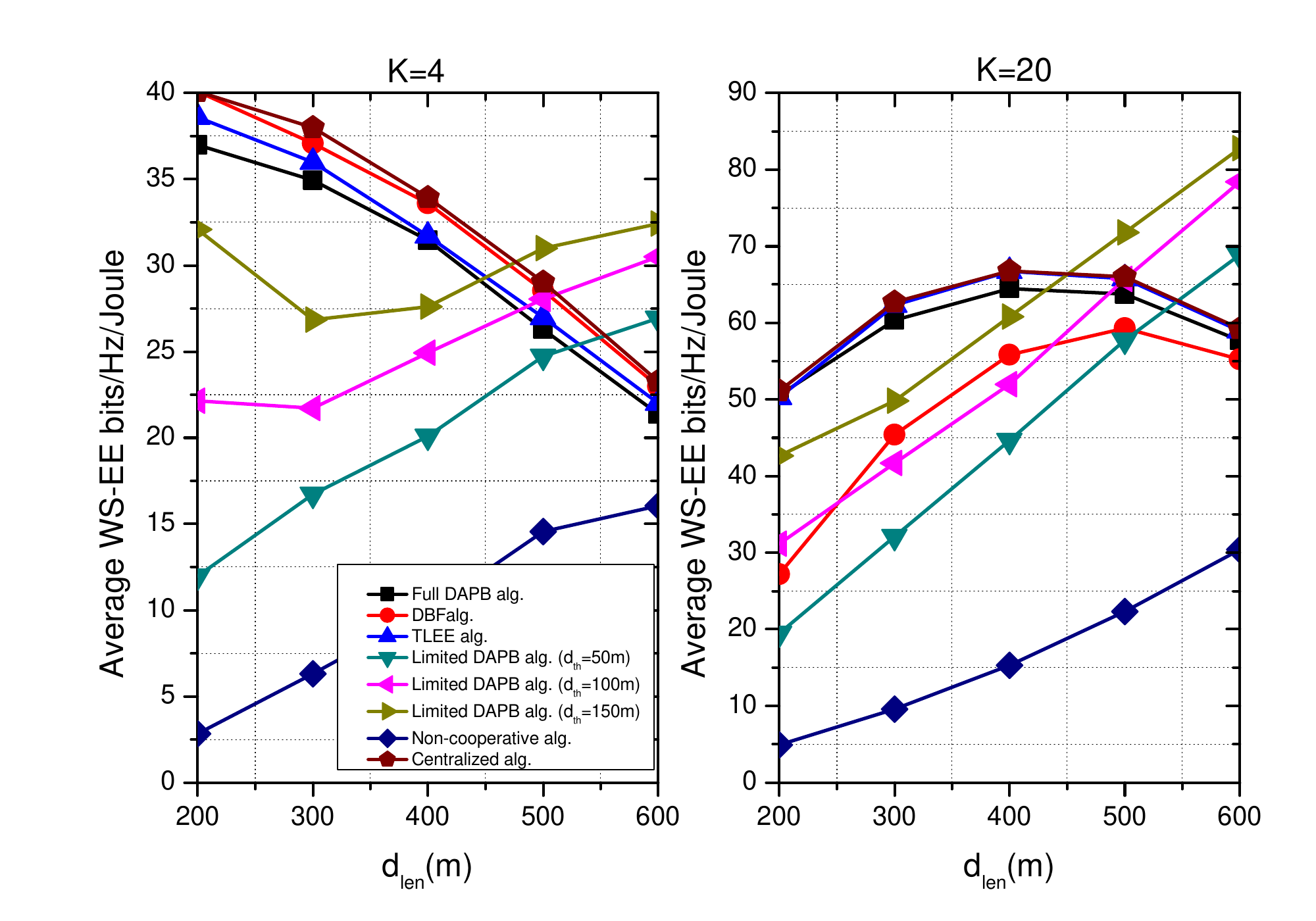}
\caption{Average WS-EE versus length $d_{\rm{len}}$ under various algorithms for (a) $K=4$  and (b) $K=20$.}
\label{EE-dlen}
\end{figure}

Fig.~\ref{EE-dlen} illustrates the average WS-EE versus length $d_{\rm{len}}$ under various algorithms for $K=4$ and $20$. The figure shows that in the sparse network with $K=4$, the WS-EE achieved by the full DAPB algorithm decreases with the increase of the length $d_{\rm{len}}$. This is due to the fact that with the increase of $d_{\rm{len}}$, the transmission distance for each user increases, leading to a higher backhaul power consumption. However, in the dense network with $K=20$, the WS-EE achieved by the full DAPB algorithm first increases and then decreases with  $d_{\rm{len}}$. The reason is that in the latter scenario,  interference is a dominant factor for limiting the system performance, and sightly increasing $d_{\rm{len}}$ will reduce the interference among users, and thus the WS-EE will increase. However, further increasing $d_{\rm{len}}$ will incur significant backhaul power consumption, resulting in the decrease of the WS-EE. It is also observed that the WS-EE achieved by the limited DAPB algorithm increases with $d_{\rm{len}}$. The reason is that increasing $d_{\rm{len}}$ will not affect the backhaul power consumption of the limited DAPB algorithm (with fixed transmission distance), but decrease the interference among users.

\begin{table}[!t]
\renewcommand{\arraystretch}{1.1}
\caption{Average number of exchanged real scalars}
\label{tab5}
\centering
\begin{tabular}{|c|c|c|c|c|c|}
\hline
  &$K=5$ &$K=20$& & $K=5$& $K=20$\\
\hline
Centralized& 240 &3360& Full DAP &128& 3950\\
\hline
$d_{\rm{th}}=50{\rm{m}}$& 24.1 &358& $d_{\rm{th}}=100{\rm{m}}$&71.2 &1350\\
\hline
$d_{\rm{th}}=150{\rm{m}}$&101.2 &1926 & Noncoop& 34.1& 111.6\\
\hline
DBF& 166.8 &3019 &TLEE &40716& 50312\\
\hline
\end{tabular}
\end{table}

Finally, we compare the feedback overhead for the above algorithms. Table \ref{tab5} shows the average number of exchanged real scalars  with $P_{\rm{max}}=35 \ {\rm{dBm}}$. Several interesting observations can be found from this table. The TLEE algorithm has much higher overhead than the other methods including the centralized algorithm, since it involves two layers of iterative procedures. Besides the requirement of a CPU by the centralized algorithm with high computational complexity, the centralized algorithm has almost twice the overhead of the full DAPB algorithm in the case of $K=5$. For a dense network with K=20, we find the similar overhead of the full DAPB algorithm as the centralized one. However, the overhead is further significantly reduced by the limited DAPB and the simulation results in Table \ref{tab5} also verify a much lower complexity of the proposed algorithm compared to the centralized one.
Interestingly, for the case of $K=5$, the overhead of the limited DAPB algorithm with $d_{\rm{th}}=50{\rm{m}}$ is even smaller than that of the non-cooperative method, the reason might be that the limited DAPB algorithm converges much faster than the non-cooperative method in this scenario. It is seen that the overhead  of the DBF algorithm is slightly smaller than that of the full DAPB algorithm. However, its computational complexity is much higher than the full DAPB algorithm as seen from Table \ref{tab4}. More importantly, its performance is usually worse than the full DAPB algorithm since it is mainly designed for symmetrical networks. Note that the limited  DAPB algorithm with $d_{\rm{th}}=150{\rm{m}}$ usually performs  better than the DBF algorithm in terms of the WS-EE as seen above, but has less information exchange overhead.

\section{Conclusion}\label{con}
In this paper, we proposed a distributed pricing-based algorithm named DAPB algorithm to address the problem of maximizing the WS-EE in MISO ICs. In this algorithm, each user updates its own beam-vector with low-complexity: the beam direction is expressed in a closed form, and the globally optimal power allocation can be obtained with simple arithmetic operations. We provided a rigorous proof for the convergence of the proposed algorithm.  Our proposed algorithm converges much faster than the DBF and TLEE algorithms. The convergence speed of DBF algorithm is significantly affected by the number of the links, while that of the TLEE algorithm is considerably affected by the maximum transmit power. The WS-EE performance is comparable to that of the centralized algorithm and the TLEE algorithm, and much better than the DBF algorithm and the noncooperative algorithm. However, our proposed algorithm has much lower computational complexity and communication overhead. It is interesting to find out that in some scenarios the limited DAPB performs better than the full DAPB algorithm due to its lower backhaul power consumption.

\numberwithin{equation}{section}
\begin{appendices}

\section{Proof of Theorem 1}
The proof is established by showing that when one user updates its beam-vector by solving problem (\ref{peruser}), the WS-EE is non-decreasing. To this end, we first show that ${U_j}({\pmb{W}})$  is convex w.r.t. the interference  ${I_j}$. The second partial derivative of ${U_j}({\pmb{W}})$  w.r.t.  ${I_j}$ can be computed as
	 	
\begin{equation}\label{secpartialderivative}
 \frac{{{\partial ^2}{U_j}({\pmb{W}})}}{{\partial I_j^2}} = \frac{{{1}}}{{{P_{j,T}}\ln 2}}\left( {A_j^{ - 2} - B_j^{ - 2}} \right) \geq 0,
\end{equation}
where ${A_j} = \sigma _j^2{\rm{ + }}{I_j}$  and ${B_j} = \sigma _j^2{\rm{ + }}{I_j} + {\left| {{\pmb{h}}_{j,j}^{\rm{H}}{{\pmb{w}}_j}} \right|^2}$. Hence, ${U_j}({\pmb{W}})$  is a convex function of  ${I_j}$.

Let ${\pmb{W}} = \left[ {{{\pmb{w}}_1}, \cdots ,{{\pmb{w}}_K}} \right]$  denote the beam-vectors of all users before user $k$   starts to update its beam-vector. Let  ${{\pmb{\tilde w}}_k}$ denote the updated beam-vector of user  $k$ with given  ${{\pmb{W}}_{ - k}}$. Let ${\pmb{\tilde W}} = ({{\pmb{\tilde w}}_k},{{\pmb{W}}_{ - k}})$  be the beam-vectors of all users after user  $k$ updates its beam-vector. We have
 \begin{align}
&\sum\limits_{j = 1}^K {{\alpha _j}{U_j}({\pmb{\tilde W}})} \\
&{\rm{ = }}{\alpha _k}{U_k}({\pmb{\tilde W}}) + \sum\limits_{j = 1,j \ne k}^K {{\alpha _j}{U_j}({\pmb{\tilde W}})} \\
& \ge {\alpha _k}{U_k}({\pmb{\tilde W}}) + \sum\limits_{j = 1,j \ne k}^K {{\alpha _j}\left( {{U_j}({\pmb{W}}) + \frac{{\partial {U_j}({\pmb{W}})}}{{\partial {I_j}}}\left( {{{\tilde I}_j} - {I_j}} \right)} \right)} \label{firstinequ}\\
&= {\alpha _k}{U_k}({\pmb{\tilde W}}) + \sum\limits_{j = 1,j \ne k}^K {{\alpha _j}{U_j}({\pmb{W}})}  - \sum\limits_{j = 1,j \ne k}^K {{\pi _j}\left( {{{\tilde I}_j} - {I_j}} \right)} \label{secondineq}\\
&= {\alpha _k}{U_k}({\pmb{\tilde W}}) + \sum\limits_{j = 1,j \ne k}^K {{\alpha _j}{U_j}({\pmb{W}})}  - {\pmb{\tilde w}}_k^{\rm{H}}{{\pmb{L}}_k}{{\pmb{\tilde w}}_k} + {\pmb{w}}_k^{\rm{H}}{{\pmb{L}}_k}{{\pmb{w}}_k}\\
& \ge {\alpha _k}{U_k}({\pmb{W}}) + \sum\limits_{j = 1,j \ne k}^K {{\alpha _j}{U_j}({\pmb{W}})}  - {\pmb{w}}_k^{\rm{H}}{{\pmb{L}}_k}{{\pmb{w}}_k} + {\pmb{w}}_k^{\rm{H}}{{\pmb{L}}_k}{{\pmb{w}}_k}\label{20}\\
&= \sum\limits_{j = 1}^K {{\alpha _j}{U_j}({\pmb{W}})}
\end{align}
where ${\tilde I_j}$  and  ${I_j}$ denote the interferences at ${\pmb{\tilde W}}$  and  ${\pmb{W}}$, respectively, (\ref{firstinequ})  is due to the convexity of function  ${U_j}({\pmb{W}})$, (\ref{secondineq}) follows by using the definition of  ${\pi _j}$, and (\ref{20}) holds because of the step 2 of the DAPB algorithm. Thus, the WS-EE is non-decreasing after each user updates its beam-vector  ${{\pmb{w}}_k}$. Furthermore, the WS-EE is upper bounded. Hence, the DAPB algorithm  converges.

\section{Proof of Theorem 2}
The first derivative of  $\phi ({p_k})$  w.r.t.  $p_k$  can be computed as
 \begin{align}
\phi '({p_k}) &= \frac{{\frac{{{g_{kk}}({p_k} + {P_{C,k}})}}{{1 + {{\tilde g}_{kk}}{p_k}}} - \ln (1 + {g_{kk}}{p_k}) - {A_k}{{({p_k} + {P_{C,k}})}^2}}}{{{{({p_k} + {P_{C,k}})}^2}}}\\
 &\buildrel \Delta \over = \frac{{\psi ({p_k})}}{{{{({p_k} + {P_{C,k}})}^2}}}.
\end{align}
The first derivative of $\psi ({p_k})$ w.r.t. $p_k$ can be calculated as
\begin{equation}\label{secderivative}
\psi '({p_k}) =  - \frac{{g_{kk}^2{P_{C,k}} + g_{kk}^2{p_k} + 2{A_k}({p_k} + {P_{C,k}}){{(1 + {g_{kk}}{p_k})}^2}}}{{{{(1 + {g_{kk}}{p_k})}^2}}}.
\end{equation}
We have $\psi '({p_k}) < 0$  for  ${\rm{0}} \le {p_k} \le {\bar P_k}$. Then the optimum solution will fall into three cases:
\begin{enumerate}
  \item If  $\phi '(0) \le 0$, then $\phi '(p_k) \le 0$ holds for ${\rm{0}} \le {p_k} \le {\bar P_k}$, implying that  $\phi ({p_k})$ is monotonically decreasing for  ${\rm{0}} \le {p_k} \le {\bar P_k}$. Hence, the optimum solution of problem (\ref{powerallo}) is  ${\mathord{\buildrel{\lower3pt\hbox{$\scriptscriptstyle\frown$}}
\over p} _k} = 0$;
  \item If  $\phi '({\bar P_k}) \ge 0$, then $\phi '(p_k) \ge 0$ holds in the feasible power region of link  $k$. Then  $\phi ({p_k})$ is monotonically increasing for  ${\rm{0}} \le {p_k} \le {\bar P_k}$. In this case, the optimum solution is  ${\mathord{\buildrel{\lower3pt\hbox{$\scriptscriptstyle\frown$}}
\over p} _k} = {\bar P_k}$;
  \item If $\phi ' (0) < 0$  and  $\phi '({\bar P_k}) > 0$, then there must exist a unique globally optimal power allocation  ${\mathord{\buildrel{\lower3pt\hbox{$\scriptscriptstyle\frown$}}
\over p} _k}$ such that  $\phi '({\mathord{\buildrel{\lower3pt\hbox{$\scriptscriptstyle\frown$}}
\over p} _k}) = 0$ due to the fact that  $\phi '({p_k})$ is monotonically decreasing for  ${\rm{0}} \le {p_k} \le {\bar P_k}$.
\end{enumerate}

\section{Proof of Theorem 3}
The globally optimal solution to problem (\ref{alternative}) can be obtained by solving the KKT conditions of problem (\ref{alternative}). The Lagrangian function of problem (\ref{alternative}) can be written as
\begin{equation}\label{Lagrangian}
{\cal L}({p_{k,1}},{p_{k,2}},\alpha ,\beta ,\gamma ) = \frac{{\ln (1 + {g_{k,1}}{p_{k,1}} + {g_{k,2}}{p_{k,2}})}}{{{p_{k,1}} + {p_{k,2}} + {P_{{C,k}}}}} - {g_{k,3}}{p_{k,1}} + \alpha {p_{k,1}} + \beta {p_{k,2}} + \gamma ({P_k} - {p_{k,1}} - {p_{k,2}}),
\end{equation}
where $\alpha ,\beta ,\gamma$ are the non-negative dual variables associated with the corresponding constraints of problem (\ref{alternative}). According to \cite{Boyd2004,bertsekas1999nonlinear}, all locally optimal solutions (including the globally optimal solution) should satisfy the KKT conditions of problem (\ref{alternative}):

\begin{eqnarray}
  - \frac{{\ln (1 + {g_{k,1}}{p_{k,1}}+ {g_{k,2}}{p_{k,2}})}}{{{{({p_{k,1}} + {p_{k,2}} + {P_{{C,k}}})}^2}}} + \frac{{{g_{k,1}}}}{{({p_{k,1}} \!+\! {p_{k,2}}\! +\! {P_{{C,k}}})(1\! +\! {g_{k,1}}{p_{k,1}}\! +\! {g_{k,2}}{p_{k,2}})}}\! -\! {g_{k,3}}\! +\! \alpha \! -\! \gamma  \!\!\!\!&=& \!\!\!\! 0,\label{derivativetheta1}\\
  - \frac{{\ln (1 \!+ \! {g_{k,1}}{p_{k,1}} \!+\! {g_{k,2}}{p_{k,2}})}}{{{{({p_{k,1}} \!+\! {p_{k,2}} \!+\! {P_{{C,k}}})}^2}}}\! +\! \frac{{{g_{k,2}}}}{{({p_{k,1}}\! +\! {p_{k,2}}\! +\! {P_{{C,k}}})(1 + {g_{k,1}}{p_{k,1}} \!+\! {g_{k,2}}{p_{k,2}})}} \!+ \!\beta  \!-\! \gamma \!\!\! \!&=&\!\!\!\! 0,\label{derivativetheta2}\\
\alpha {p_{k,1}} \!\!\!\!&=&\!\!\!\! 0,\label{afatheta}\\
\beta {p_{k,2}} \!\!\!\!&=&\!\!\!\! 0,\label{betatheta}\\
\gamma ({P_k} - {p_{k,1}} - {p_{k,2}})\!\!\!\! &=&\!\!\!\! 0,\label{powercomple}\\
\alpha ,\beta ,\gamma  \ge 0,{p_{k,1}},{p_{k,2}} \ge 0,{p_{k,1}} + {p_{k,2}} \!\!\!\!&\le& \!\!\!\!{P_k}.\label{feasibitheta}
\end{eqnarray}
Our main idea is to obtain all feasible solutions to the above KKT conditions and pick the best one as the final solution, which is definitely globally optimal. According to (\ref{feasibitheta}), the optimal solution must fall into seven cases:  (1) ${p_{k,1}} = 0$, ${P_k} > {p_{k,2}} > 0$; (2) ${p_{k,1}} > 0$, ${p_{k,2}} > 0$, ${p_{k,1}} + {p_{k,2}} < {P_k}$; (3) ${p_{k,1}} > 0$, ${p_{k,2}} > 0$, ${p_{k,1}} + {p_{k,2}} = {P_k}$; (4) ${p_{k,1}} = 0$, $ {p_{k,2}}={P_k} $; (5) ${P_k} > {p_{k,1}} > 0$, ${p_{k,2}} = 0$; (6) ${p_{k,1}}={P_k} $, ${p_{k,2}} = 0$; (7) ${p_{k,1}} = 0$, ${p_{k,2}} = 0$.
In the following, we characterize the conditions for each individual case to be satisfied.
\begin{enumerate}

  \item \textbf{Case} ${p_{k,1}} = 0$, ${P_k} > {p_{k,2}} > 0$: From (\ref{betatheta}) and (\ref{powercomple}), it follows that $\beta  = \gamma  = 0$. By applying these facts to (\ref{derivativetheta2}), ${p_{k,2}}$ should satisfy the equality $f_1({p_{k,2}})=0$. The derivative of $f_1({p_{k,2}})$ w.r.t. ${p_{k,2}}$ is $f_1'({p_{k,2}}) =  - {g_{k,2}}\ln (1 + {g_{k,2}}{p_{k,2}}) < 0$. Hence, $f_1({p_{k,2}})$ is decreasing in ${p_{k,2}}$. To find one feasible ${p_{k,2}} \in (0,{P_k})$, two conditions should be satisfied:
      \begin{eqnarray}
          f_1(0) \!\!&=& \!\!{g_{k,2}}{P_{{C,k}}} > 0,\\
          f_1({P_k})\!\!\! &<&\!\!\!0\label{secondeq}.
     \end{eqnarray}
     The first condition follows obviously, hence  only the second condition needs to be checked. If condition (\ref{secondeq}) is satisfied, there exists a unique solution to equation $f_1({p_{k,2}})=0$. Rearranging the terms, we have
      \begin{eqnarray}
 f_1({p_{k,2}})\!\!&=&\!\!- ({g_{k,2}}{p_{k,2}} + 1){\rm{ln}}({g_{k,2}}{p_{k,2}} + 1) + ({g_{k,2}}{p_{k,2}} + 1) - (1 - {g_{k,2}}{P_{C,k}})\\
 \!\!&=&\!\!  - e({\rm{ln}}({g_{k,2}}{p_{k,2}} + 1) - 1){\rm{exp}}\{ {\rm{ln}}({g_{k,2}}{p_{k,2}} + 1) - 1\}  - (1 - {g_{k,2}}{P_{C,k}})\\
 \!\! &=&\!\! 0.
  \end{eqnarray}
The closed-form solution is given by (\ref{pk2op}). The final task is to verify whether condition (\ref{derivativetheta1}) can be satisfied. By inserting ${p_{k,2}^\star}$ from (\ref{pk2op}) into (\ref{derivativetheta1}) and using the facts that $\gamma =0$ and ${p_{k,1}}= 0$, we have $\alpha$ in (\ref{afta}). If $\alpha\geq0$,  one locally optimal solution is given by ${p_{k,1}} = 0$, ${p_{k,2}} = {p_{k,2}^\star}$.

  \item \textbf{Case} ${p_{k,1}} > 0,{p_{k,2}} > 0$, ${p_{k,1}} + {p_{k,2}} < {P_k}$: From (\ref{afatheta})-(\ref{powercomple}), it follows that $\alpha  = \beta  = \gamma  = 0$. Hence, equations (\ref{derivativetheta1}) and (\ref{derivativetheta2}) reduce to
       \begin{eqnarray}
          \!- \frac{{\ln (1 \!+\! {g_{k,1}}{p_{k,1}} + {g_{k,2}}{p_{k,2}})}}{{{{({p_{k,1}} \!+\! {p_{k,2}} + {P_{{C,k}}})}^2}}} \!\!+\!\! \frac{{{g_{k,1}}}}{{({p_{k,1}} + {p_{k,2}}\! +\! {P_{{C,k}}})(1 \!+\! {g_{k,1}}{p_{k,1}}\! +\! {g_{k,2}}{p_{k,2}})}} - {g_{k,3}}\!\!\!\! &=&\!\!\!\! 0,\label{222uppperequa}\\
          - \frac{{\ln (1 + {g_{k,1}}{p_{k,1}} + {g_{k,2}}{p_{k,2}})}}{{{{({p_{k,1}} + {p_{k,2}} + {P_{{C,k}}})}^2}}} + \frac{{{g_{k,2}}}}{{({p_{k,1}} + {p_{k,2}} + {P_{{C,k}}})(1 + {g_{k,1}}{p_{k,1}} + {g_{k,2}}{p_{k,2}})}}\!\!\!\! \!&=&\!\!\!\!\! 0,\label{222downequ}
     \end{eqnarray}
     (\ref{222uppperequa})  minus (\ref{222downequ}) yields
     \begin{equation}\label{f3}
      ({p_{k,1}} + {p_{k,2}} + {P_{{C,k}}})(1 + {g_{k,1}}{p_{k,1}} + {g_{k,2}}{p_{k,2}}) = \frac{{{g_{k,1}} - {g_{k,2}}}}{{{g_{k,3}}}}.
     \end{equation}
     Hence, ${{g_{k,1}} > {g_{k,2}}}$ must hold. Obviously, $\frac{{{g_{k,1}} - {g_{k,2}}}}{{{g_{k,3}}}}$ must be larger than ${{P_{{C,k}}}}$ (when ${{p_{k,1}}}$ and ${{p_{k,2}}}$ approach zero), and be smaller than $({P_k} + {P_{{C,k}}})(1 + {g_{k,1}}{P_k})$ (when ${{p_{k,1}}}$ approaches ${P_k}$ and ${{p_{k,2}}}$ approaches zero). Hence, the chain of inequalities in  (\ref{inequallll}) should hold. If it is true, define ${\theta  \triangleq({p_{k,1}} + {p_{k,2}} + {P_{{C,k}}})}$, and from (\ref{f3}) we have $(1 + {g_{k,1}}{p_{k,1}} + {g_{k,2}}{p_{k,2}}) = \frac{{{g_{k,1}} - {g_{k,2}}}}{{{g_{k,3}}}}\frac{1}{\theta }$. Using these definitions, (\ref{222downequ}) can be transformed into ${f_3}(\theta )=0$.
     The derivative of  ${f_3}(\theta )$ w.r.t. $\theta $ is given by
     \begin{equation}\label{derivativeoff3}
       {{f'}_3}(\theta ) = \frac{1}{\theta } + \frac{{2{g_{k,2}}{g_{k,3}}}}{{{g_{k,1}} - {g_{k,2}}}}\theta  > 0.
     \end{equation}
     This means that ${f_3}(\theta )$ is a monotonically increasing function of $\theta$. Since ${P_k} + {P_{{C,k}}} > \theta  > {P_{{C,k}}}$, two conditions should be satisfied to ensure that  ${f_3}(\theta )=0$ has a solution:
     ${f_3}({P_{{C,k}}}) < 0$ and ${f_3}({P_k} + {P_{C,k}}) > 0$. If both conditions are satisfied, employ the bisection search method to find the solution to equation ${f_3}(\theta )=0$, denoted as $\theta^\star $. Then, $\theta_1$ and $\theta_2$ should satisfy the following set of equalities
     \begin{equation}\label{setofequ}
     \left\{ \begin{array}{l}
     1 + {g_{k,1}}{p_{k,1}} + {g_{k,2}}{p_{k,2}}{\rm{ = }}\frac{{{g_{k,1}} - {g_{k,2}}}}{{{g_{k,3}}}}\frac{1}{\theta^\star }\\
     {p_{k,1}} + {p_{k,2}} + {P_{{C,k}}}{\rm{ = }}\theta^\star.
      \end{array} \right.
     \end{equation}
     Hence, one locally optimal solution is given by (\ref{locallycase6}).

  \item \textbf{Case} ${p_{k,1}} > 0,{p_{k,2}} > 0$, ${p_{k,1}} + {p_{k,2}} = {P_k}$: From (\ref{afatheta}) and (\ref{betatheta}), it follows that $\alpha  = \beta  = 0$. Combining with the fact that ${p_{k,1}} + {p_{k,2}} = {P_k}$, (\ref{derivativetheta1}) and (\ref{derivativetheta2}) reduce to
      \begin{eqnarray}
          - \frac{{\ln (1 + {g_{k,1}}{p_{k,1}} + {g_{k,2}}{p_{k,2}})}}{{{{({P_k} + {P_{{C,k}}})}^2}}} + \frac{{{g_{k,1}}}}{{({P_k} + {P_{{C,k}}})(1 + {g_{k,1}}{p_{k,1}} + {g_{k,2}}{p_{k,2}})}} - {g_{k,3}} - \gamma  \!\!&=&\!\! 0,\label{upper}\\
          - \frac{{\ln (1 + {g_{k,1}}{p_{k,1}} + {g_{k,2}}{p_{k,2}})}}{{{{({P_k} + {P_{{C,k}}})}^2}}} + \frac{{{g_{k,2}}}}{{({P_k} + {P_{{C,k}}})(1 + {g_{k,1}}{p_{k,1}} + {g_{k,2}}{p_{k,2}})}} - \gamma \!\!&=&\!\! 0,\label{lower}
     \end{eqnarray}
      (\ref{upper}) minus (\ref{lower}) yielding
      \begin{equation}\label{dddd}
        1 + {g_{k,1}}{p_{k,1}} + {g_{k,2}}{p_{k,2}} = \frac{{{g_{k,1}} - {g_{k,2}}}}{{{g_{k,3}}({P_k} + {P_{{C,k}}})}}.
      \end{equation}
      Hence, ${g_{k,1}} > {g_{k,2}}$ must hold.  If ${g_{k,1}} > {g_{k,2}}$,  $\gamma $ can be solved from (\ref{lower})  as in (\ref{gama}). If $\gamma \geq 0$, combining (\ref{dddd}) with the fact that ${p_{k,1}} + {p_{k,2}} = {P_k}$, we can obtain one locally optimal solution as in (\ref{last}).
\end{enumerate}
Following analogous  analysis, the conditions for the other cases follow immediately and we can prove that Case 7) never happens. The details are omitted.

\section{Proof of Lemma 1}
We first prove that if $g_{k,1}\leq g_{k,2}$,  the optimal $p_{k,1}$ must be zero. This can be proved by contradiction. Assume $\{p_{k,1}^{(1)}>0, p_{k,2}^{(1)}\geq 0\}$ is the optimal solution. One can construct another feasible solution: $\{p_{k,1}^{(2)}=0, p_{k,2}^{(2)}=p_{k,2}^{(1)}+p_{k,1}^{(1)}\}$ that leads to a larger objective value in (\ref{alternative}), which contradicts the assumption. Then, problem (\ref{alternative}) reduces to the following optimization problem
 \begin{equation}\label{simplifiedoptimization}
\begin{array}{l}
\mathop {\max }\limits_{0 \le {p_{k,2}} \le {P_k}} f({p_{k,2}}) = \frac{{\ln (1 + {g_{k,2}}{p_{k,2}})}}{{{p_{k,2}} + {P_{C,k}}}}.
\end{array}
\end{equation}
Note that this problem is a special case of problem (\ref{powerallo}) when $A_k=0$. Hence, it can be solved by the method in Theorem 2. However, for this specific form of the problem, we can fortunately obtain a closed-form solution. Similar to the analysis in Theorem 2, we can easily prove that function $f(\cdot)$ first increases and then decreases in the region ${p_{k,2}} \in [0,\infty )$. Hence, there exists a unique maximum solution $p_{k,2}^\star$ that satisfies the first-order optimal condition $f'({p_{k,2}}) = 0$, which can be equivalently expressed as
\begin{equation}\label{equality}
{g_{k,2}}(p_{k,2}^* + {P_{C,k}}) - (1 + {g_{k,2}}p_{k,2}^*)\ln (1 + {g_{k,2}}p_{k,2}^\star) = 0.
\end{equation}
Employing the similar analysis to Case 1) in Theorem 3, we obtain the optimal solution $p_{k,2}^\star$ in (\ref{simplifiedsolution}).

\section{Centralized algorithm}
Due to the non-concavity of the overall objective function in (\ref{mainproblem}), it is difficult to obtain a globally optimal solution. Inspired by the centralized algorithm in \cite{Sigen2003}, we employ the gradient projection method  combined with the Armijo rule to find a local optimum. Let ${{\cal W}_k} = \left\{ {{{\pmb{w}}_k}|{{\pmb{w}}_k} \in {\mathbb{C}^{M \times 1}},{{\left\| {{{\pmb{w}}_k}} \right\|}^2} \le {P_k}} \right\}$ as the feasible beamforming set for user $k$, which is a convex set. Define  ${P_{{\cal W}_k}}\{\cdot \} $ as the projection onto  ${{\cal W}_k}$. Then the centralized algorithm is given in Algorithm 2,  where  ${U_{{\rm{ws}}}}({\pmb{W}}) =\sum\limits_{k = 1}^K {{\alpha _k}{U_k}({\pmb{W}})}$. To make the algorithm  work, there are still two problems to be solved: how to calculate  the gradient and how to perform the projection.

\begin{itemize}
  \item	Gradient calculation: The partial derivative of ${U_{{\rm{ws}}}}({\pmb{W}})$  w.r.t. ${{\pmb{w}}_k}$  is given by
      \begin{equation}\label{derivative}
       {\nabla _{{{\pmb{w}}_k}}}{U_{{\rm{ws}}}}({\pmb{W}}) = \frac{{{\alpha _k}}}{{\ln 2}}\frac{{{P_{k,T}}{{\pmb{h}}_{k,k}}{\pmb{h}}_{k,k}^{\rm{H}}{{\pmb{w}}_k} - \rho \ln (1 + {\eta _k})(\sigma _k^2 + {I_k}){{\pmb{w}}_k}}}{{P_{k,T}^2(\sigma _k^2 + {I_k} + |{\pmb{h}}_{k,k}^{\rm{H}}{{\pmb{w}}_k}{|^2})}} - {{\pmb{L}}_k}{{\pmb{w}}_k}¡£
      \end{equation}
  \item Projection: The projection of ${{\pmb{w}}_k}$ onto the convex set ${{\cal W}_k}$  is given by
  \begin{equation}\label{projection}
    {P_{{{\cal W}_k}}}\{ {{\pmb{w}}_k}\}  = \left\{ \begin{array}{l}
\sqrt {{P_k}} \frac{{{{\pmb{w}}_k}}}{{\left\| {{{\pmb{w}}_k}} \right\|}},\ {\rm{  if  }}\left\| {{{\pmb{w}}_k}} \right\|^2 > {P_k}{\rm{, }}\\
{{\pmb{w}}_k},\quad\qquad {\rm{              Otherwise}}{\rm{.}}
\end{array} \right.
  \end{equation}
\end{itemize}
\begin{algorithm}[H]
\caption{Centralized EE Maximization Algorithm}
\begin{algorithmic}[1]
\STATE Initialize any feasible  ${{\pmb{W}}^{(0)}} = [{\pmb{w}}_1^{(0)},{\pmb{w}}_2^{(0)}, \cdots ,{\pmb{w}}_K^{(0)}]$, tolerance $\varepsilon $ , iteration number  $n = 0$, $\delta  = 0.3$,  $\beta  = 0.5$.
\STATE	Calculate the gradient  ${\pmb{g}}_k^{(n)} = {\nabla _{{{\pmb{w}}_k}}}{U_{{\rm{ws}}}}({\pmb{W}}),\forall k$;
\STATE Choose appropriate ${s^{(n)}}$  and let ${\pmb{\bar w}}_k^{(n)} = {P_{{{\cal W}_k}}}\{ {\pmb{w}}_k^{(n)} + {s^{(n)}}{\pmb{g}}_k^{(n)}\}, \forall k $;
\STATE Update  ${\pmb{w}}_k^{(n + 1)} = {\pmb{w}}_k^{(n)} + {\kappa ^{(n)}}({\pmb{\bar w}}_k^{(n)} - {\pmb{w}}_k^{(n)}),\forall k$, where ${\kappa  ^{(n)}} = {\beta ^{{m^{(n)}}}}$  and  ${m^{(n)}}$ is the first non-negative integer $m$  that satisfies
\[{U_{{\rm{ws}}}}({{\pmb{W}}^{(n + 1)}}) - {U_{{\rm{ws}}}}({{\pmb{W}}^{(n)}}) \ge \delta {\beta ^m}\sum\limits_{k = 1}^K {{\pmb{g}}_k^{(n){\rm{H}}}({\pmb{\bar w}}_k^{(n)} - {\pmb{w}}_k^{(n)})}. \]
\STATE Let ${{\pmb{W}}^{(n)}} = [{\pmb{w}}_1^{(n)},{\pmb{w}}_2^{(n)}, \cdots ,{\pmb{w}}_K^{(n)}]$.  If  ${{\left| {{U_{{\rm{ws}}}}({{\pmb{W}}^{(n + 1)}}) - {U_{{\rm{ws}}}}({{\pmb{W}}^{(n)}})} \right|} \mathord{\left/
 {\vphantom {{\left| {{U_{{\rm{ws}}}}({{\pmb{W}}^{(n + 1)}}) - {U_{{\rm{ws}}}}({{\pmb{W}}^{(n)}})} \right|} {{U_{{\rm{ws}}}}({{\pmb{W}}^{(n)}}) < \varepsilon }}} \right.
 \kern-\nulldelimiterspace} {{U_{{\rm{ws}}}}({{\pmb{W}}^{(n)}}) < \varepsilon }}$, terminate.  Otherwise, set $n \leftarrow n + 1$  and go to step 2.
\end{algorithmic}
\end{algorithm}

\end{appendices}




\
\



\bibliographystyle{IEEEtran}
\bibliography{myre}



\end{document}